\documentclass[journal]{IEEEtran}
\usepackage{amsmath,amsfonts}
\usepackage{algorithmic}
\usepackage{algorithm}
\usepackage{array}
\usepackage[caption=false,font=footnotesize,labelfont=rm,textfont=rm]{subfig}
\usepackage{textcomp}
\usepackage{stfloats}
\usepackage{url}
\usepackage{verbatim}
\usepackage{graphicx}
\usepackage{cite}

\usepackage{hyperref}
\usepackage{makecell}
\usepackage{multicol}
\usepackage{multirow}
\usepackage{bm}
\usepackage{diagbox}
\usepackage{amssymb}
\usepackage{threeparttable}
\usepackage{amsthm}
\usepackage{color}

\theoremstyle{definition}

\DeclareSubrefFormat{myparens}{#1(#2)} 
\makeatletter
\renewcommand{\citepunct}{,\penalty\@m\hskip.13emplus.1emminus.1em}
\renewcommand{\citedash}{\hbox{--}\penalty\@m}

\usepackage[utf8]{inputenc}
\usepackage[T1]{fontenc}

\begin{document}

\title{Learning Wideband User Scheduling and Hybrid Precoding with Graph Neural Networks}

\author{Shengjie Liu, Chenyang Yang,~\IEEEmembership{Senior Member,~IEEE}, and Shengqian Han,~\IEEEmembership{Senior Member,~IEEE}


\thanks{A part of this work has been presented in IEEE WCNC 2024 \cite{LSJ_WCNC}.}

\thanks{The authors are with the School of Electronic and Information Engineering, Beihang University, Beijing 100191, China (e-mail: liushengjie@buaa.edu.cn; cyyang@buaa.edu.cn; sqhan@buaa.edu.cn).}

\thanks{The source code is available at \href{https://github.com/LSJ-BUAA/GNN-Scheduling-Precoding}{https://github.com/LSJ-BUAA/GNN-Scheduling-Precoding}.}

}

\markboth{Journal of \LaTeX\ Class Files,~Vol.~0, No.~0, July~2025}%
{Shell \MakeLowercase{\textit{et al.}}: A Sample Article Using IEEEtran.cls for IEEE Journals}


\maketitle

\begin{abstract}
User scheduling and hybrid precoding in wideband multi-antenna systems have never been learned jointly due to the challenges arising from the massive user combinations on resource blocks (RBs) and the shared analog precoder among RBs. In this paper, we strive to jointly learn the scheduling and precoding policies with graph neural networks (GNNs), which have emerged as a powerful tool for optimizing resource allocation thanks to their potential in generalizing across problem scales. By reformulating the joint optimization problem into an equivalent functional optimization problem for the scheduling and precoding policies, we propose a GNN-based architecture consisting of two cascaded modules to learn the two policies. We discover a same-parameter same-decision (SPSD) property for wireless policies defined on sets, revealing that a GNN cannot well learn the optimal scheduling policy when users have similar channels. This motivates us to develop a sequence of GNNs to enhance the scheduler module. Furthermore, by analyzing the SPSD property, we find when linear aggregators in GNNs impede size generalization. Based on the observation, we devise a novel attention mechanism for information aggregation in the precoder module. Simulation results demonstrate that the proposed architecture achieves satisfactory spectral efficiency with short inference time and low training complexity, and is generalizable to the numbers of users, RBs, and antennas at the base station and users.
\end{abstract}


\section{Introduction}\label{sec:intro}
\IEEEPARstart{H}{ybrid} analog and baseband precoding stands as a pivotal technique for supporting high spectral efficiency (SE) in millimeter wave  multiple-input multiple-output (MIMO) systems \cite{overviewHY}. While the decreased number of radio frequency (RF) chains reduces the hardware and energy costs, it also imposes a constraint on the number of users served on each resource block (RB) in orthogonal frequency division multiplexing (OFDM) systems.

Wideband MIMO systems require optimizing spatial-frequency user scheduling, i.e., deciding which users should be spatially multiplexed on which RBs. The scheduling is further coupled with hybrid precoding, making their optimization highly challenging. Specifically, the analog precoder, shared among all RBs \cite{MO}, is affected by the scheduling decisions across RBs, thereby impeding the application of existing narrow-band user scheduling methods on each RB. The computational complexity of exhaustively searching all possible user combinations grows exponentially with the numbers of candidate users and RBs \cite{tmlcn}. To avoid the prohibitive complexity of jointly optimizing spatial-frequency scheduling and hybrid precoding, several heuristic methods were proposed, say imposing zero-forcing~(ZF) constraint on precoding in linear successive allocation (LISA) \cite{LISA}, combining greedy scheduling with ZF precoder \cite{Bogale}, or ignoring the multi-user interference (MUI) during scheduling \cite{KLX}. However, these methods still suffer from high computational costs or lead to evident performance degradation.

To facilitate real-time implementation, learning-based methods have been developed. Wireless policies, which map environment parameters such as channels into resource allocation outcomes, can be learned by deep neural networks (DNNs) with short inference time. Yet, to the best of our knowledge, the study of jointly learning spatial-frequency user scheduling and hybrid precoding has never been reported.

\subsection{Related Works}
\subsubsection{Learning precoding and scheduling policies}
Most previous works studied the learning of precoding and scheduling in narrow-band systems. A majority of these studies learned to optimize scheduling with a pre-determined precoder, where the DNNs were trained in a supervised manner \cite{FNNsupervised,SPAWCsupervised} or by reinforcement learning (RL) \cite{RL-SLNR,RLschedule2} to deal with the difficulty of learning ``0-1'' variables. In \cite{FNNsupervised}, a fully-connected neural network (FNN) was applied to learn a scheduling policy given maximum ratio transmission. In \cite{SPAWCsupervised}, Transformer \cite{transformer} was used to select users with an existing approach for learning combinatorial problems, where ZF precoder was considered.
In \cite{RL-SLNR}, an FNN was trained by RL to optimize scheduling, given a baseband precoder that maximizes signal-to-leakage-and-noise ratio. In \cite{RLschedule2}, RL was also adopted for user scheduling, given an analog precoder comprising the eigenvectors of channel covariance matrix and a ZF baseband precoder.

Several recent works attempted to learn both scheduling and precoding policies in narrow-band systems. In \cite{codebook}, an FNN integrated with attention mechanism was applied to schedule users and choose analog precoder from a codebook, where the baseband precoding was solved with an existing numerical algorithm. In \cite{yuweiSchedule}, two graph neural networks (GNNs), respectively designed for user scheduling and reconfigurable intelligent surface (RIS) configuration, were trained separately with unsupervised learning, and then the weighted minimum mean square error (WMMSE) algorithm was used for optimizing precoding.
In \cite{RLschedule1}, a RIS configuration and baseband precoder module was trained by an RL algorithm, then the module was fixed for training a scheduler module with another RL algorithm, and the two modules were not trained jointly.
To jointly optimize user scheduling and baseband precoding in narrow-band multi-user multiple-input single-output (MU-MISO) systems, a GNN was designed in \cite{HSW}, where the structure of the optimal precoder matrix was harnessed to simplify the functions to be learned. However, the method is problem-specific and not applicable to wideband or hybrid precoding systems.

For wideband systems, the learning of scheduling was studied in \cite{tmlcn}, where two FNNs were employed as actor and critic networks. Although analog precoder was not involved, the scheduling problem is coupled across RBs due to the objective of user fairness. However, a ZF precoder was used and the joint learning was not investigated in \cite{tmlcn}.


\subsubsection{Learning to optimize with GNNs} GNNs have been shown to surpass FNNs and convolutional neural networks (CNNs) in terms of training efficiency and the potential in size generalizability \cite{LYreview}. GNNs, exploiting the permutation prior of wireless policies, such as permutation equivariance (PE) and permutation invariance (PI), learn policies in smaller hypothesis spaces, thereby reducing training complexity. Once trained, GNNs can ``be applied to'' systems of varying scales because their update equations are independent of graph sizes \cite{LYreview}. However, the generalization performance of GNNs is not guaranteed. A GNN is ``generalizable to'' problem scales only if it can be well-performed as the scales change. Several factors, including policies to be learned, types of sizes considered, and the design of GNNs, affect the generalizability. It can be observed from simulation results in the literature that the size generalizability of GNNs differs among different policies and types of sizes. Taking precoding policy as an example, GNNs need to be judiciously designed for being generalized well to the number of users \cite{modelGNN}, whereas GNNs using linear aggregator can achieve favorable generalization performance to the number of antennas \cite{LSJ_TWC,ZBC_TCOM}.


\subsubsection{Attention mechanism for precoding}
In order to improve the learning performance and size generalizability, attention mechanism has been introduced for optimizing precoding. In \cite{dll,RLschedule1}, the attention mechanism in Transformer \cite{transformer} was employed to capture MUI and boost SE. In \cite{GATbeijiao,GATbeijiao2}, graph attention network (GAT) \cite{GAT} was used to learn over a graph with only user vertices to enable generalizability to the number of users. In \cite{liyang}, Transformer was regarded as a GNN, which was shown to be generalizable to the number of users.
In \cite{LSJ_TWC, modelGNN}, attention mechanisms were designed for the edge-GNNs, where edge representations were updated to facilitate generalizability to the number of users. Notwithstanding, all aforementioned works consider narrow-band MU-MISO systems. The designed attention mechanisms are not applicable to wideband MU-MIMO systems.

\subsection{Motivation and Contributions}


Although existing works studied the learning of scheduling and precoding, the joint learning of spatial-frequency scheduling, hybrid precoding, and analog combining remains unsolved. This problem involves learning multiple high-dimensional decisions, necessitating efficient DNN designs.

Previous studies have recognized that GNN is a powerful tool for optimizing precoding, with the potential in generalizing across problem scales, particularly when associated with attention mechanisms.
However, incorporating an attention mechanism is not always beneficial in generalizability while adding computation complexity. 
Unfortunately, existing studies lack the foresight to identify which types of sizes GNNs struggle to generalize well without attention mechanisms. As a result, extensive trial-and-error procedures are required for the design of GNNs. Therefore, it is crucial to determine which types of information should be aggregated using attention mechanisms when designing GNNs.

In this paper, we propose a GNN-based architecture to jointly learn spatial-frequency scheduling, hybrid precoding, and analog combining in downlink wideband MU-MIMO systems. Simulation results are provided to evaluate the proposed architecture in terms of learning performance, size generalizability, training complexity, and inference time by comparing it with numerical algorithms and other DNNs.

The major contributions are summarized as follows.
\begin{itemize}
\item We transform the joint optimization problem into a functional optimization problem for a two-layer nested policy, consisting of an inner scheduling policy (corresponding to spatial-frequency scheduling) and an outer precoding policy (corresponding to hybrid precoding and analog combining). To learn the two policies, we propose an architecture comprising a scheduler module and a precoder module, which can be jointly trained in an unsupervised manner to enhance learning performance. Both modules are designed as GNNs that satisfy the complex permutation properties of the two policies, thereby reducing training complexity and achieving size generalizability.

\item To provide guidance for designing GNNs with good size generalizability and learning performance, we discover a same-parameter same-decision (SPSD) property for the optimization problems and resulting policies defined on sets. This property impacts GNNs in two ways. First, we observe that GNNs with linear aggregators can be generalized to the size of a set when the policy to be learned is SPSD on the set, making attention mechanism unnecessary. Second, we find a mismatch between the functions learned by GNNs and non-SPSD policies, resulting in a challenge for GNNs in learning these policies when some environment parameters are similar.

\item We find that the precoding policy is non-SPSD on the user set, leading to poor generalizability to the number of users for GNNs with linear aggregators. To address this, we devise a partial attention mechanism for the GNN in the precoder module, which can distinguish the importance of different users with multiple antennas on every RB during information aggregation.

\item We find that the scheduling policy is non-SPSD on the user set, causing a single GNN to be unable to well learn the scheduling policy when users have similar channels. To tackle this, we propose a sequence of GNNs as an enhancement of the scheduler module to improve the learning performance in high-density user scenarios.
\end{itemize}

Compared to the conference version \cite{LSJ_WCNC}, this journal version finds the SPSD property, extends the problem from narrow-band user scheduling and hybrid precoding to wideband spatial-frequency scheduling, hybrid precoding and combining, develops a novel GNN architecture with reduced training and inference complexities, and provides more simulation results for both performance and complexity comparisons.

The rest of the paper is organized as follows. Sec.~\ref{sec:SPSD} introduces the SPSD property and its impacts on GNNs. Sec.~\ref{sec:model} introduces the joint scheduling and precoding problem and the two policies to be learned. Sec.~\ref{sec:GNN} proposes our architecture and analyzes its computational complexity. Sec.~\ref{sec:sim} provides simulation results. Sec.~\ref{sec:con} provides conclusions.

\emph{Notations:} $\odot$ denotes the element-wise product of two vectors. $\otimes$ denotes the Kronecker product of two matrices. $\bf \Pi$ denotes permutation matrix. $\pi(\cdot)$ maps the $i$th element in a set into the $\pi(i)$th element. $({\bf X})_{i_1,\cdots,i_N}$ denotes an element with index $i_1,\cdots,i_N$ in an $N$-dimensional array $\bf X$.


\section{SPSD Property of Wireless Policies}\label{sec:SPSD}
In this section, we introduce the SPSD property of wireless policies defined on sets and discuss the impacts of the property.

\subsection{SPSD Property}
A wireless policy is a mapping from the environment parameters to the optimal solutions of an optimization problem.
Wireless policies obtained from the problems consisting of sets are with permutation properties \cite{LSJ_TWC}. For instance, consider the following optimization problem
\begin{subequations}
\begin{align}
        &P_1: \max_{{\bf w}_1,\cdots,{\bf w}_N} ~~ F_0({\bf w}_1,\cdots,{\bf w}_N,{\bf h}_1,\cdots,{\bf h}_N),\nonumber\\
        &{\rm s.t.} ~ F_i({\bf w}_1,\cdots,{\bf w}_N,{\bf h}_1,\cdots,{\bf h}_N)\leq 0, i\!=\!1,\cdots,N_C,
\end{align}
\end{subequations}
where $F_0(\cdot)$ is the objective function, $F_i(\cdot)$ is the $i$th constraint function, and $N$ is the problem scale.

The policy obtained from $P_1$ is the mapping from the environment parameters ${\bf h}_1,\cdots,{\bf h}_N$ to the corresponding optimal decisions ${\bf w}_1,\cdots,{\bf w}_N$, denoted as $({\bf w}_1,\cdots,{\bf w}_N) = f({\bf h}_1,\cdots,{\bf h}_N)$. The policy $f(\cdot)$ is a multivariate function if $P_1$ has a unique optimal solution. It is not a function if multiple optimal solutions exist, since a function is a one-to-one or many-to-one~mapping.

$P_1$ involves a set of size $N$, which is $\{1,\cdots,N\}\triangleq \mathcal{N}$. If permuting the indices $\{1,\cdots,N\}$ into $[\pi(1),\cdots,\pi(N)]^T = {\bf \Pi}^T[1,\cdots,N]^T$ does not change the objective and constraint functions of $P_1$, then the policy $f(\cdot)$ satisfies the \emph{PE property}, i.e.,  $({\bf w}_{\pi(1)},\cdots,{\bf w}_{\pi(N)}) = f({\bf h}_{\pi(1)},\cdots,{\bf h}_{\pi(N)})$ \cite{LSJ_TWC}, and $f(\cdot)$ is called a PE-policy.

The decisions ${\bf w}_1,\cdots,{\bf w}_N$ are the optimal solution corresponding to a group of parameters ${\bf h}_1,\cdots,{\bf h}_N$. If ${\bf h}_i={\bf h}_j$, then ${\bf w}_i={\bf w}_j$ or ${\bf w}_i\neq{\bf w}_j$ depending on the problem.

\noindent\underline{\textbf{Definition 1:} \emph{(PE-)SPSD property.}} Assume that ${\bf h}_i={\bf h}_j, i\neq j$. If the optimal solution of $P_1$ yields ${\bf w}_i={\bf w}_j$, then problem $P_1$ and the PE-policy $f(\cdot)$ are PE-SPSD or simply SPSD on the set $\mathcal{N}$. If ${\bf w}_i\neq {\bf w}_j$, then $P_1$ and $f(\cdot)$ are non-SPSD on the set $\mathcal{N}$.

To help understand the property, we provide two examples.


\noindent\underline{\textbf{Example 1:} \emph{Narrow-band MU-MISO precoding.}} 
The precoding problem for SE maximization under the transmit power constraint consists of two sets. One set is composed of $K$ users, and the other set is composed of $N_T$ antennas. The corresponding precoding policy can be expressed as $({\bf w}_1,\cdots,{\bf w}_K) = f({\bf h}_1,\cdots,{\bf h}_K)$, which is defined on user set with ${\bf w}_k,{\bf h}_k\in \mathbb{C}^{N_T\times 1}$. The policy can also be expressed as $(\tilde{\bf w}_1,\cdots,\tilde{\bf w}_{N_T}) = f(\tilde{\bf h}_1,\cdots,\tilde{\bf h}_{N_T})$, which is defined on antenna set with $\tilde{\bf w}_n,\tilde{\bf h}_n\in \mathbb{C}^{K\times 1}$. As proved in Appendix~\ref{app:A}, this problem and the policy are SPSD on antenna set, but are non-SPSD on user set.

\noindent\underline{\textbf{Example 2:} \emph{Power control in interference channels.}}~The power control problem for SE maximization in an interference system with two transceiver pairs consists of one set, i.e., the set of transceiver pairs. As proved in Appendix~\ref{app:B}, the SPSD property of this problem and the policy depends on the value of a threshold $s_h$ related to channel gains. They are SPSD for large $s_h$ but non-SPSD for small $s_h$.


\noindent\underline{\textbf{Definition 2:} \emph{PI-SPSD property.}}
Consider the policy ${\bf w} = f({\bf h}_1,\cdots,{\bf h}_N)$. If the policy exhibits the \emph{PI property} on $\mathcal{N}$, i.e., ${\bf w} = f({\bf h}_{\pi(1)},\cdots,{\bf h}_{\pi(N)})$, then the policy is PI-SPSD.

The PI-SPSD property can be regarded as a special case of PE-SPSD. To see this, one can rewrite the PI-policy ${\bf w} = f({\bf h}_1,\cdots,{\bf h}_N)$ as $({\bf w}_1,\cdots,{\bf w}_N) = f({\bf h}_1,\cdots,{\bf h}_N)$ subject to ${\bf w}_1=\cdots={\bf w}_N={\bf w}$, which is naturally SPSD.

\vspace{-4mm}\subsection{Impacts of SPSD Property on GNNs}\label{subsec:2b}\vspace{-1mm}
%
%
%
%

\subsubsection{\underline{Impact on Size Generalization}}
A GNN can satisfy the permutation property of a policy via a proper design~\cite{LSJ_TWC}. After training, the GNN can be used for different problem scales. However, the generalization performance is not guaranteed~\cite{RGNN}.




From empirical evaluations in the literature, we can observe that the size generalizability of GNNs is affected by the SPSD property of a policy. Taking the MU-MISO precoding policy (i.e., Example 1) as an instance, a GNN satisfying the PE properties on both user and antenna sets is easy to be generalized to the number of antennas~\cite{LSJ_TWC,ZBC_TCOM}.
That is, the GNN achieves satisfactory generalization performance to the number of antennas, even if the information from different antennas is simply linearly aggregated in the update process.
However, the GNN can be generalized to the number of users only if its update equation is judiciously designed~\cite{modelGNN}, since the policy is non-SPSD on user set.


\subsubsection{\underline{Impact on Learning Performance}}
A function learned by a GNN over a graph with $N$ vertices can be denoted as
$({\bf w}_1,\cdots\!,{\bf w}_i,\cdots\!,{\bf w}_j,\cdots\!,{\bf w}_N) \!=\!g({\bf h}_1,\cdots\!,{\bf h}_i,\cdots\!,{\bf h}_j,\cdots\!,{\bf h}_N)$. When the GNN satisfies the same PE property as the policy to be learned, ${\bf w}_i$ and ${\bf w}_j$ will be swapped if ${\bf h}_i$ and ${\bf h}_j$ are swapped. Thus, the function can also be expressed as $({\bf w}_1,\cdots\!,{\bf w}_j,\cdots\!,{\bf w}_i,\cdots\!,{\bf w}_N) \!=\! g({\bf h}_1,\cdots\!,{\bf h}_j,\cdots\!,{\bf h}_i,\cdots\!,{\bf h}_N)$. If ${\bf h}_i={\bf h}_j$, the two groups of inputs become identical, and the corresponding two groups of outputs will also be identical, which implies ${\bf w}_i={\bf w}_j$, since a GNN learns functions rather than mappings. This indicates that a GNN cannot learn the optimal solution of a non-SPSD problem when some of the environment parameters are identical. Even though it is unlikely for two environment parameters to be exactly the same in practice, learning a non-SPSD policy with a GNN is challenging for achieving good performance. This is because a GNN produces continuous functions, leading to ${\bf w}_i\approx {\bf w}_j$ if ${\bf h}_i\approx {\bf h}_j$.

\section{Scheduling and Precoding Policies}\label{sec:model}
In this section, we formulate the joint scheduling and precoding problem in wideband MU-MIMO systems, and transform it into a functional optimization without loss of optimality. Then, we analyze the permutation properties and SPSD properties of the scheduling and precoding policies.

\subsection{Joint Scheduling and Precoding Optimization Problem} \label{sec:model11}
Consider a downlink MU-MIMO OFDM system, where a BS equipped with $N_T$ antennas and $N_{\text{RF}}$ RF chains serves $K$ candidate users over $M$ RBs. Each RF chain is connected to all antennas \cite{MO}. Each user has $N_R$ antennas and one RF chain. At most $K'$ users can be scheduled on each RB, where $K'\leq N_{\text{RF}}$. A hybrid analog and baseband precoding is used for $K'$ scheduled users.

We first introduce some notations.
\begin{itemize}
  \item \textbf{Channels:} ${\bf H} \in \mathbb{C}^{M\times KN_R\times N_T}$ represents the wideband channels from the BS to all candidate users. ${\bf H}_m\in \mathbb{C}^{KN_R\times N_T}$ is the channel matrix of all candidate users on the $m$th RB. ${\bf H}_{m,k}\in \mathbb{C}^{N_R\times N_T}$ is the channel matrix of the $k$th user on the $m$th RB. ${\bf h}_{m,kr}\in \mathbb{C}^{N_T\times 1}$ is the channel vector to the $r$th receive antenna at the $k$th user on the $m$th RB. 
  \item \textbf{Analog precoder:} ${\bf W}_{\text{RF}}\in \mathbb{C}^{N_T\times N_{\text{RF}}}$ denotes the analog precoding matrix. 
  \item \textbf{Baseband precoder:} ${\bf W}_{\text{BB}}\in \mathbb{C}^{M\times N_{\text{RF}}\times K}$ represents the wideband baseband precoders of all candidate users. ${\bf W}_{\text{BB}m}\in \mathbb{C}^{N_{\text{RF}}\times K}$ is the baseband precoding matrix of all candidate users on the $m$th RB. ${\bf w}_{\text{BB}m,k}\in \mathbb{C}^{N_{\text{RF}}\times 1}$ is the baseband precoding vector of the $k$th user on the $m$th RB. 
  \item \textbf{Analog combiner:} ${\bf v}_{\text{RF}}\in \mathbb{C}^{KN_R\times 1}$ denotes the analog combining vector of all candidate users. ${\bf v}_{\text{RF}k}\in \mathbb{C}^{N_R\times 1}$ denotes the analog combining vector of the $k$th user. 
  \item \textbf{Scheduler decision:} ${\bf A}\in \{0,1\}^{M\times K}$ denotes the scheduling matrix. ${\bf a}_m\in \{0,1\}^{1\times K}$ denotes the scheduling decisions on the $m$th RB. $a_{m,k}\in \{0,1\}$ indicates whether the $k$th user is scheduled on the $m$th RB.
  \item \textbf{Notations for scheduled users:} we use ${\bf H}'$, ${\bf W}_{\text{BB}}'$, and ${\bf v}_{\text{RF}}'$ to denote the counterparts of the notations for $K'$ scheduled users, which have the same dimensions as ${\bf H}$, ${\bf W}_{\text{BB}}$, and ${\bf v}_{\text{RF}}$, but with $K'$ replacing $K$.
\end{itemize}

Spatial-frequency user scheduling, hybrid precoding, and analog combining can be optimized jointly, say to maximize SE as follows \cite{SVDYH, LISA}, 
\begin{subequations}
    \begin{align}
        P_J:&\hspace{-4mm} \max_{{\bf A},{\bf v}_{\text{RF}},{\bf W}_{\text{RF}},{\bf W}_{\text{BB}}} \hspace{-4mm} R({\bf A},{\bf v}_{\text{RF}},{\bf W}_{\text{RF}},{\bf W}_{\text{BB}};{\bf H}) \!\triangleq\! \frac{1}{M} \sum_{m=1}^M \sum_{k=1}^K R_{m,k}  \nonumber \\
	    {\rm s.t.} ~
        & \textstyle{\sum}_{m=1}^M \textstyle{\sum}_{k=1}^K a_{m,k}\left\|{\bf W}_{\text{RF}} {\bf w}_{\text{BB}{m,k}}\right\|_2^2 = P_{tot},\label{prob:1a}  \\
	    &|({\bf W}_{\text{RF}})_{n,j}|\!=\!1, n\!=\!1,\cdots,N_T,j\!=\!1,\cdots,N_{\text{RF}},\label{prob:1b}\\
        &|({\bf v}_{\text{RF}})_{kr}|=1, k=1,\cdots,K, r=1,\cdots,N_R,\label{prob:1c}\\
        &a_{m,k} \in \{0,1\},m=1,\cdots,M, k=1,\cdots,K,\label{prob:1d}\\
	    & \textstyle{\sum}_{k=1}^K a_{m,k} \leq K', m=1,\cdots,M,\label{prob:1e}
    \end{align}
\end{subequations}
where $R_{m,k}$ is the data rate of the $k$th user on the $m$th~RB, $
  {R_{m,k} \!=\! \log_2 \! \bigg(1+\frac{a_{m,k}|{\bf v}_{\text{RF}k}^H{\bf H}_{m,k} {\bf W}_{\text{RF}} {\bf w}_{\text{BB}{m,k}}|^2}{\sum_{\genfrac{}{}{0pt}{}{i=1}{i\neq k}}^K \! a_{m,i}|{\bf v}_{\text{RF}k}^H{\bf H}_{m,k} {\bf W}_{\text{RF}} {\bf w}_{\text{BB}{m,i}}|^2 + N_R\sigma^2}\!\bigg)}$,
and $\sigma^2$ is noise power that is amplified by ${\bf v}_{\text{RF}k}^H{\bf v}_{\text{RF}k}=N_R$.

In problem $P_J$, constraint \eqref{prob:1a} is the total power constraint, \eqref{prob:1b} and \eqref{prob:1c} are respectively the constant modulus constraints for analog precoder and analog combiners, and \eqref{prob:1e} restricts the maximal number of scheduled users on each RB. In fact, tightening constraint \eqref{prob:1e} as
\begin{equation}
     \sum_{k=1}^K a_{m,k} = K', m=1,\cdots,M, \label{prob:1e'}
\end{equation}
does not affect the optimality. This is because both $a_{m,k}=0$ and ${\bf w}_{\text{BB}{m,k}}={\bf 0}$ indicate not to schedule the $k$th user on the $m$th RB. In other words, the optimal baseband precoder will allocate zero power to the unscheduled users, even if their indicators are ``1''s.

The motivation for jointly optimizing analog combining, hybrid precoding, and scheduling is to maximize the achievable sum rate. Optimizing these components separately or neglecting any of them will lead to performance~loss. While the joint optimization requires that the analog combiners are computed at the BS and subsequently informed to the users, the associated overhead is not large. This is because the analog combiners, which consist only of phase shifters, are shared across all RBs. Consequently, only $N_R$ phases within a single analog combiner need to be sent to each scheduled user, irrespective of the number of RBs.

%

\subsection{Problem Reformulation and Two Policies to Be Learned}

According to the proof in \cite{Sun2019Learning}, the parameter optimization problem $P_J$ can be equivalently transformed into a functional optimization problem with respect to the joint policy $({\bf A},{\bf v}_{\text{RF}},{\bf W}_{\text{RF}},{\bf W}_{\text{BB}})=f_J({\bf H})$, which can be formulated as
\begin{align*}
    P_F: \max_{f_J(\cdot)} ~~&\mathbb{E}\left\{ R(f_J({\bf H});{\bf H}) \right\} \\
                {\rm s.t.} \  ~~   & \eqref{prob:1a} \sim \eqref{prob:1d}, \eqref{prob:1e'},
\end{align*}	
where $f_J(\cdot)$ maps channels $\mathbf{H}$ to the optimal solution of problem $P_J$, and the expectation $\mathbb{E}\{\cdot\}$ is taken over the random channels $\mathbf{H}$.

Learning-based methods employ DNNs to parameterize the policy $f_J(\cdot)$. However, directly learning the joint policy $f_J(\cdot)$ as in \cite{LSJ_WCNC} entails an excessive number of invalid decisions, such as the precoding and combining vectors for unscheduled users. This significantly increases learning complexity.

To avoid learning invalid decisions, we transform the joint policy into a two-layer nested policy. Specifically, the inner layer is the \emph{scheduling policy}, denoted by ${\bf A} = f_S({\bf H})$, while the outer layer is the \emph{precoding policy} learning only for the scheduled users, denoted by $({\bf v}'_{\text{RF}},{\bf W}_{\text{RF}},{\bf W}'_{\text{BB}}) = f_P({\bf H}')$. The channels of scheduled users ${\bf H}'$ is selected from ${\bf H}$ based on the value of ${\bf A}$ and can be obtained as
\begin{equation}
   {\bf H}' = f_H({\bf H},{\bf A})\in \mathbb{C}^{M\times K'N_R\times N_T}, \label{eq:fH}
\end{equation}
where the function $f_H(\cdot)$ will be detailed in the next section.
Upon substituting $f_S(\cdot)$ into \eqref{eq:fH} and then into $f_P(\cdot)$, we can obtain the two-layer nested policy~as
\begin{equation}
    ({\bf v}'_{\text{RF}},{\bf W}_{\text{RF}},{\bf W}'_{\text{BB}}) = f_P(f_H({\bf H},f_S({\bf H}))).\label{eq:nested}
\end{equation}

Considering that the unscheduled users contribute zero data rate, the objective of problem $P_J$ is equal~to
\begin{equation}
     R'\left({\bf v}'_{\text{RF}},{\bf W}_{\text{RF}},{\bf W}'_{\text{BB}}; {\bf H}' \right) \! \triangleq \! \frac{1}{M} \!\!\sum_{m=1}^M \!\sum_{k'=1}^{K'}\!\! R'_{m,k'},
\end{equation}
where $R'_{m,k'} \!=\! \log_2 \! \bigg(\!1\!+\!\frac{|{\bf v}_{\text{RF}{k'}}^{'H}{\bf H}'_{m,k'} {\bf W}_{\text{RF}} {\bf w}'_{\text{BB}{m,k'}}|^2}{\sum_{\genfrac{}{}{0pt}{}{i=1~}{i\neq k'}}^{K'} \! |{\bf v}_{\text{RF}{k'}}^{'H}{\bf H}'_{m,k'} {\bf W}_{\text{RF}} {\bf w}'_{\text{BB}{m,i}}|^2 + N_R\sigma^2}\!\bigg)$. Consequently, $P_F$ can be equivalently transformed into the following functional optimization problem with respect to $f_P(f_H(\cdot,f_S(\cdot)))$,
\begin{subequations}
    \begin{align}
        &P_N:  \hspace{-3mm} \max_{f_P(f_H(\cdot,f_S(\cdot)))} \hspace{-3mm} \mathbb{E}\left\{ R'\left(f_P(f_H({\bf H},f_S({\bf H}))) ; f_H({\bf H},f_S({\bf H})) \right) \right\} \nonumber \\
        &~~{\rm s.t.}~~\sum_{m=1}^M \sum_{k'=1}^{K'} \left\|{\bf W}_{\text{RF}} {\bf w}'_{\text{BB}{m,k'}}\right\|_2^2 = P_{tot},\label{prob:pa}  \\
        &~~~~~~~~|({\bf v}'_{\text{RF}})_{k'r}|=1, k'=1,\cdots,K', r=1,\cdots,N_R,\label{prob:pc}\\
        & ~~~~~~~~  \eqref{prob:1b}, \eqref{prob:1d}, \eqref{prob:1e'}. \nonumber 
    \end{align}
\end{subequations}
Note that, problem $P_N$ and $P_F$ have equivalent constraints ($P_N$ omits those for unscheduled users) and have equal objective function as aforementioned. Thus, problem $P_N$ is equivalent to $P_F$ in the sense that they can achieve the same optimal objective value.

From \eqref{eq:nested}, we can observe that the input to the outer policy $f_P(\cdot)$ depends on the inner policy $f_S(\cdot)$, indicating that $f_P(\cdot)$ has to be jointly optimized with~$f_S(\cdot)$. To achieve this, we propose a cascaded architecture to jointly learn the two policies, where a \emph{scheduler module} learns $f_S(\cdot)$ and a \emph{precoder module} learns $f_P(\cdot)$. The output of the scheduler module is passed to $f_H(\cdot)$, and the resulting output is then passed to the precoder module. Both modules are parameterized as DNNs and require joint training to optimize their performance.

\subsection{Permutation Properties of the Two Policies}\label{subsec:3c}
To enhance training efficiency and enable size generalization, we design GNN to realize the two modules in the proposed architecture. The designed GNNs should satisfy the same permutation properties as the scheduling and precoding policies. We analyze the permutation properties of both policies in the following.

According to the method given in \cite{LSJ_TWC}, the permutation properties can be analyzed by identifying sets relevant to a policy. We can find three sets related to the two policies: RB, BS antenna (AN$^{\rm BS}$), and user antenna (AN$^{\rm UE}$). The RB set comprises $M$ RBs. The AN$^{\rm BS}$ set comprises $N_T$ antennas at the BS.
The AN$^{\rm UE}$ set is a nested set, consisting of subsets (each corresponding to a user). Specifically, for scheduling, the AN$^{\rm UE}$ set comprises $K$ subsets, each containing $N_R$ antennas. For precoding, the AN$^{\rm UE}$ set comprises $K'$ subsets, each containing $N_R$ antennas. For brevity, we refer to the whole AN$^{\rm UE}$ set as a \emph{user set} with $K$ (or $K'$) ``elements'', each ``element'' is a subset, denoted as an \emph{AN} set.



In Table \ref{table:property}, we summarize the permutation properties of the two policies with respect to each of these sets, instead of expressing the properties using the permutation matrices as in \cite{LSJ_TWC}. To aid in understanding the table, we elaborate on the first two rows in the ``RB'' column: a) For the scheduling policy, if the indices of RBs are permuted in ${\bf H}$, the indices of RBs in scheduling decision ${\bf A}$ will be permuted equivariantly. b) If the indices of RBs are permuted in ${\bf H}'$, the precoding decision ${\bf v}'_{\text{RF}}$ will remain unchanged (i.e., invariant), since the analog combiner is shared among all RBs.

\begin{table}[htb!]
    \setlength\tabcolsep{1.5pt}
\centering
\vspace{-0mm}
\caption{Permutation and SPSD Properties of the Two Policies}\label{table:property}
\vspace{-2mm}
\footnotesize
    \begin{tabular}{c|c|c|c|c|c}
        \hline\hline
            \multicolumn{2}{c|}{\multirow{2}{*}{\diagbox[width=22mm]{\bf Policy}{\bf Set}}} & \multirow{2}{*}{RB} & \multirow{2}{*}{AN$^{\rm BS}$} & \multicolumn{2}{c}{AN$^{\rm UE}$ (nested)} \\ \cline{5-6}
            \multicolumn{2}{c|}{} &  &  & \multicolumn{1}{c|}{User} & AN \\ \hline
            Scheduling                  & $\bf A$           & PE SPSD & PI SPSD & PE SPSD/non-SPSD & PI SPSD \\\hline
            \multirow{3}{*}{Precoding}  & ${\bf v}'_{\text{RF}}$   & PI SPSD & PI SPSD & PE non-SPSD & PE SPSD \\\cline{2-6}
                                        & ${\bf W}_{\text{RF}}$    & PI SPSD & PE SPSD & PI SPSD & PI SPSD \\\cline{2-6}
                                        & ${\bf W}'_{\text{BB}}$   & PE SPSD & PI SPSD & PE non-SPSD & PI SPSD \\\hline
            \hline
    \end{tabular}
    \vspace{-4mm}
\end{table}

\subsection{SPSD Properties of the Two Policies}\label{subsec:3d}

Formally proving whether a PE policy is SPSD is generally challenging, especially for intractable optimization problems, e.g., the non-convex and non-differentiable problem $P_J$.
To find the SPSD properties of the scheduling and precoding policies, we adopt an empirical approach, which serves as a practical means for identifying such properties. Specifically, we generated 1,000 random channel samples using the channel model to be introduced in Sec.~\ref{sec:sim} with various sizes ($M,K,N_R,N_T$). Furthermore, to ensure the reliability of our observations and avoid conclusions biased by a single algorithm, we employ multiple algorithms for both scheduling (i.e, LISA~\cite{LISA} and Multicarrier semi-orthogonal user selection (SUS)~\cite{MSUS}) and precoding (i.e., semidefinite relaxation (SDR)~\cite{KLX}, manifold optimization (MO)~\cite{MO}, and singular value decomposition (SVD)~\cite{SVDYH}).

To infer the SPSD properties on the RB set, for instance, we randomly duplicate one channel matrix across two RBs for each sample. We then input these samples into different algorithms. For all channel samples, we consistently observe that both scheduling algorithms produce identical scheduling decisions for the two RBs, and all precoding algorithms also yield identical baseband precoders for the two RBs. These results provide empirical evidence that the scheduling and precoding policies are SPSD on the RB set. The same procedure can be applied to find SPSD properties on other~sets.

The SPSD properties of the two policies are also presented in Table \ref{table:property}. The scheduling policy on  user set may exhibit SPSD or non-SPSD property, depending on the channels. For instance, for the two users having the same channel, if the channel is weak, then neither will be scheduled, resulting in an SPSD property. Conversely, if the channel is strong, only one user will be scheduled, leading to a non-SPSD property.

As discussed in Sec.~\ref{subsec:2b}, the impact of the SPSD property is two-fold. 

Regarding generalization performance, the GNNs in both modules can be easily generalized to the numbers of RBs ($M$), antennas at the BS ($N_T$), and antennas at each user ($N_R$), because both policies are SPSD on these sets. The GNN in the scheduler module can also be easily generalized to the number of users ($K$), since the scheduling policy tends to be SPSD on user set for most samples. By contrast, the GNN in the precoder module requires careful design to be generalized to $K$ (when $K<K'$). If $K\geq K'$, the precoder module always serves $K'$ users on each RB and does not need to be generalized beyond this number.

Regarding learning performance, the GNN in the scheduler module faces challenges when learning the scheduling policy in scenario with densely distributed users. In such scenario, two users are likely to have similar channels, and the GNN schedules both or neither of them, leading to a mismatch with the non-SPSD scheduling policy and degradation in scheduling performance.
On the other hand, although the precoding policy is always non-SPSD on the user set, the performance of the GNN in the precoder module does not degrade because no users have similar channels after the scheduling.

\section{GNN-based Architecture for Joint Learning}\label{sec:GNN}
In this section, we propose the GNN-based architecture of DNN, including the scheduler and precoder modules. To facilitate joint training of the two modules, we design a differentiable $f_H(\cdot)$ that connects the two modules. Then, we construct graphs and design GNNs for the two modules. To improve the scheduling performance in crowded user scenarios, we proceed to design a sequence of GNNs for the scheduler module. To allow size generalizability, we design a partial attention mechanism for the precoder module. Finally, we introduce the training and testing procedures and analyze the inference complexity.

\begin{figure}[htb!]
	\centering
	\vspace{-0mm}
	\includegraphics[width=1.0\linewidth]{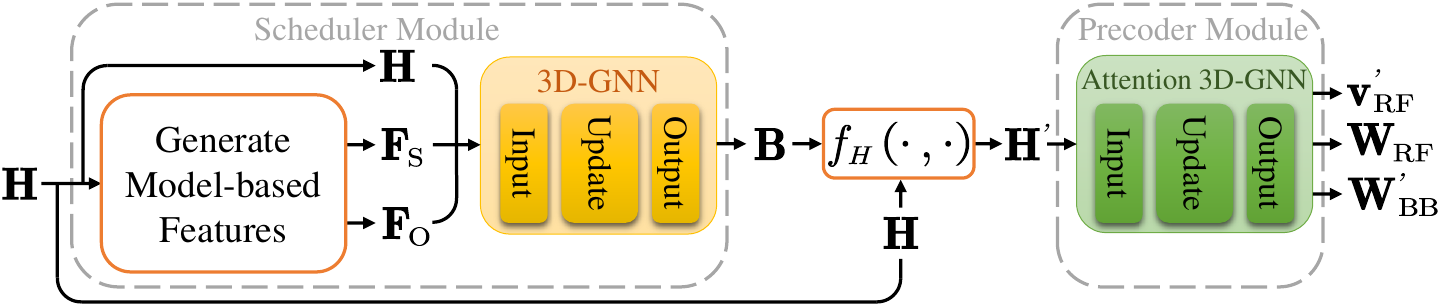}
	\vspace{-6mm}
	\caption{NGNN architecture. The GNN in scheduler module is a 3D-GNN, and the GNN in precoder module is a 3D-GNN with attention mechanism.}
        \label{fig:arch}
	\vspace{-4mm}
\end{figure}

The scheduler module can be realized as either a single GNN or a sequence of GNNs as an enhancement. This leads to two variants of the whole architecture, which are respectively referred to as non-sequential GNN (NGNN) and sequential GNN (SGNN). The NGNN architecture is shown in Fig.~\ref{fig:arch}.

\subsection{Design of $f_H(\cdot)$}

To learn $f_S(\cdot)$ and $f_P(\cdot)$, the loss function can be chosen as the negative objective function of problem $P_N$. To back-propagate the gradients of the loss function to the trainable weights in the scheduler module, the binary scheduling matrix ${\bf A}$ involved in $f_H(\cdot)$ as defined in \eqref{eq:fH} needs to be relaxed as a continuous variable during the training phase, and the function $f_H(\cdot)$ should be differentiable. 


To this end, we realize $f_H(\cdot)$ by using basis vectors to extract the channels of scheduled users.
Specifically, the matrix ${\bf A}\in \{0,1\}^{M\times K}$ is replaced by ${\bf B}\in \{0,1\}^{M\times K' \times K}$, whose $m$th slice is ${\bf B}_m \in \{0,1\}^{K' \times K}$. ${\bf B}_m$ comprises $K'$ different (row) basis vectors ${\bf b}_{m,k'}\in \{0,1\}^{1\times K}, k'=1,\cdots,K'$, satisfying $\sum_{k'=1}^{K'}{\bf b}_{m,k'}={\bf a}_m$.
There is a single ``1'' and $K-1$ ``0''s in each vector ${\bf b}_{m,k'}$. $b_{m,k',k}\in \{0,1\}$ denotes the $k$th element in ${\bf b}_{m,k'}$.


Consequently, $f_H(\cdot)$ can be expressed as a slice-wise function for each RB. Specifically, the channels of the scheduled users on the $m$th RB, ${\bf H}'_m$, can be extracted by multiplying the basis vectors with ${\bf H}_m$ as ${\bf H}'_m = ({\bf B}_m\otimes {\bf I}_{N_R}) {\bf H}_m \in \mathbb{C}^{K'N_R\times N_T}$, which is the explicit expression of $f_H(\cdot)$ on the~RB.

\subsection{Graph Construction}

Graph construction is crucial for the learning efficiency and size generalizability of GNNs. The graphs need to be designed according to the permutation properties of the two polices~\cite{LSJ_TWC}, to ensure that GNNs satisfy the matched permutation properties. 
Following the design principles from \cite{LSJ_TWC}, the number of vertex types should equal the number of sets because the unordered elements within each set can be regarded as permutable vertices of the same type. Hence, three types of vertices are defined corresponding to the three sets listed in Table \ref{table:property}, including RB, AN$^{\rm BS}$, and AN$^{\rm UE}$ vertices, where the $N_R$ AN$^{\rm UE}$ vertices at each user are referred to as a group of vertices since they belong to one subset.



\begin{figure}[!htb]
    \vspace{-3mm}
    \centering
    \hspace{-4mm}
    \subfloat[Scheduling graph. $K=3$.]{
        \begin{minipage}[c]{0.5\linewidth}
            \label{fig:graph-schedule}
            \centering
            \includegraphics[width=1.05\linewidth]{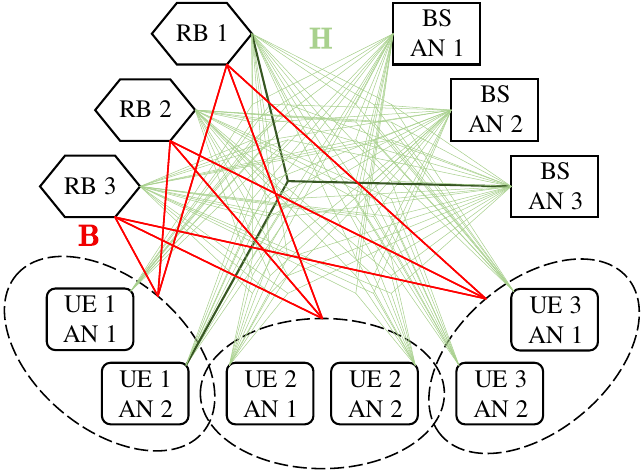}
            \vspace{-5mm}
        \end{minipage}
    }
    \hspace{-8mm}
    \subfloat[Precoding graph. $K'=2$.]{
        \begin{minipage}[c]{0.5\linewidth}
            \label{fig:graph-precode}
            \centering
            \includegraphics[width=1.05\linewidth]{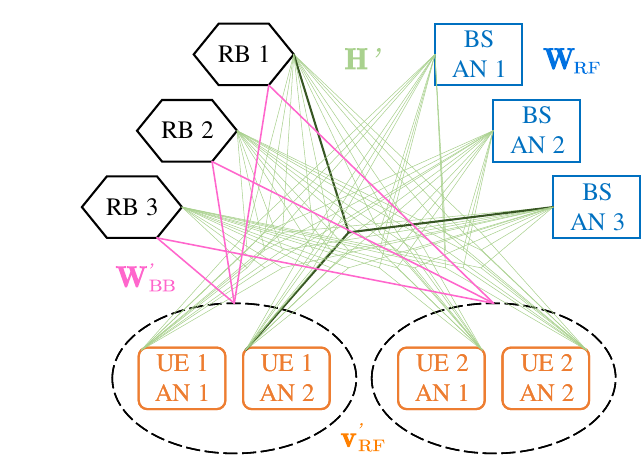}
            \vspace{-5mm}
        \end{minipage}
    }
    \vspace{-0mm}
    \caption{Two constructed graphs. $M=3, N_R=2, N_T=3$. Each hyper-edge connects three vertices of three types (shown in green). In (a), we highlight the hyper-edge connecting the 1st RB, the 2nd AN$^{\rm UE}$ of the 1st candidate user, and the 3rd AN$^{\rm BS}$ vertices (in dark green). The feature of this hyper-edge is $({\bf H})_{1,12,3}$. In (b), we highlight the hyper-edge connecting the 1st RB, the 2nd AN$^{\rm UE}$ of the 1st scheduled user, and the 3rd AN$^{\rm BS}$ vertices (in dark green), with a feature of $({\bf H}')_{1,12,3}$.}
    \label{fig:graph}
    \vspace{-2mm}
\end{figure}

\emph{Features} are the input of GNNs. In both graphs, each element in the channel array ${\bf H}$ or ${\bf H}'$ is defined as a feature of a \emph{hyper-edge} connecting one RB, one AN$^{\rm BS}$, and one AN$^{\rm UE}$ vertices, as illustrated in green in Fig. \ref{fig:graph}.
For example, $({\bf H})_{m,kr,n}$ is the feature of the hyper-edge connecting the $m$th RB, the $n$th AN$^{\rm BS}$, and the $kr$th AN$^{\rm UE}$ vertices in Fig.~\subref*{fig:graph-schedule}.

To improve the learning performance of the scheduling policy, we introduce two extra features in addition to $\mathbf{H}$ for the scheduling graph. 

First, since scheduling largely depends on the strength of channels, the first model-based feature is defined as the channel strength 
\begin{equation}
  \bar{F}_{\text{S}m,k} = ||{\bf H}_{m,k}||_F \in \mathbb{R},\ \forall m,k.
\end{equation}
This feature is associated with the edge connecting an RB vertex and a group of AN$^{\rm UE}$ vertices.

Second, the SUS algorithm \cite{SUS} suggests the importance of orthogonality between the channels of two users on each RB for scheduling. Considering the multiple antennas at each user, the second model-based feature is the averaged correlation coefficient between one channel vector and all others on an RB
\begin{equation}
\bar{F}_{\text{O}{m,kr}} \!=\! \frac{1}{K\!N_R}\!\sum_{i=1}^K\!\sum_{u=1}^{N_R}\!\frac{|{\bf h}_{m,iu}^H{\bf h}_{m,kr}|}{||{\bf h}_{m,iu}||_2||{\bf h}_{m,kr}||_2\!} \!\in\! \mathbb{R}, \!\forall m,k,r.
\end{equation}
This feature is associated with the edge connecting an RB vertex and an AN$^{\rm UE}$ vertex. 

To enhance generalization performance and mitigate the impact of varying distributions of the two model-based features with respect to the number of transmit antennas $N_T$, receive antennas $N_R$, and candidate users $K$, we normalize the features as
\begin{align}
  {F}_{\text{S}m,k} &= \frac{\bar{F}_{\text{S}m,k}-{\rm mean}_{i\in\{1,\cdots,K\}}(\bar{F}_{\text{S}m,i})}{{\rm std}_{i\in \{1,\cdots,K\}}(\bar{F}_{\text{S}m,i)}}, \\
  {F}_{\text{O}{m,kr}} &= \frac{\bar{F}_{\text{O}{m,kr}}-{\rm mean}_{iu\in\{11,\cdots,KN_R\}}(\bar{F}_{\text{O}{m,iu}})}{{\rm std}_{iu\in \{11,\cdots,KN_R\}}(\bar{F}_{\text{O}{m,iu}})}.
\end{align}

The two model-based features are not presented in Fig.~\subref*{fig:graph-schedule} for the sake of clarity.

\emph{Actions} are the output of GNNs. An action is defined on vertices or edges. Specifically, in the scheduling graph, as illustrated in Fig. \subref*{fig:graph-schedule}, each basis vector in ${\bf B}$ represents an action on an edge connecting an RB vertex and a group of AN$^{\rm UE}$ vertices (shown in red).
In the precoding graph, as illustrated in Fig. \subref*{fig:graph-precode}, each element in ${\bf v}'_{\text{RF}}$ is defined as an action on AN$^{\rm UE}$ vertex (shown in orange), each vector in ${\bf W}_{\text{RF}}$ is defined as an action on AN$^{\rm BS}$ vertex (shown in blue), and each vector in ${\bf W}'_{\text{BB}}$ is defined as an action on an edge between an RB vertex and a group of AN$^{\rm UE}$ vertices (shown in pink).

The sizes of the constructed graphs change with problem scales. Specifically, the numbers of RB, AN$^{\rm BS}$, AN$^{\rm UE}$ vertices in a group, and AN$^{\rm UE}$ groups, i.e., $M$, $N_T$, $N_R$, and $K$, vary with problem scales. Note that the value of $K'$ is constrained by $N_{\rm RF}$, meaning that the precoding graph always has $K'$ AN$^{\rm UE}$ groups (corresponding to $K'$ scheduled users as specified in \eqref{prob:1e'}), unless $K < K'$ where only $K$ AN$^{\rm UE}$ groups exist.

\subsection{Design of Scheduler Module}\label{subsec:scheModule}

In this subsection, based on the constructed graph, we design a 3D-GNN for the scheduler module. 
We first design the three key processes for the 3D-GNN, and then enhance the GNN to learn the non-SPSD scheduling policy in the scenario with densely distributed users.

In the 3D-GNN, similar to the features ${\bf H}$, hidden representations are also defined on ``hyper-edges'', which explains the ``3D'' designation \cite{LSJ_TWC}. In particular, ${\bf x}^{l}_{m,kr,n} \in \mathbb{R}^{C_l}$ denotes hidden representation of the hyper-edge connecting the $m$th RB, the $kr$th AN$^{\rm UE}$, and the $n$th AN$^{\rm BS}$ vertices in the $l$th layer, where $l=1,\cdots,L_S$, $L_S$ is the total number of layers of the GNN, and ${C_l}$ is the number of elements in hidden representation of the $l$th layer.


The 3D-GNN comprises three key processes: input, update, and output \cite{LSJ_TWC}. The input process obtains initial representations in the first layer from the features. The update process iteratively updates hidden representations by aggregating the representations from adjacent hyper-edges. The output process produces actions from the hidden representations in the last layer, and ensures that these actions satisfy the constraints.

The three processes are detailed as follows.


\subsubsection{\underline{Input Process}}

This process aims to generate all $MKN_RN_T$ representations in the first layer from features. Each representation is given by ${\bf x}^{1}_{m,kr,n} \!\!=\! [{\rm Re}(({\bf H})_{m,kr,n}), {\rm Im}(({\bf H})_{m,kr,n}),$ ${F}_{\text{S}m,k}, {F}_{\text{O}{m,kr}}]^T \in \mathbb{R}^{4}$, which is associated with the corresponding hyper-edge as shown in Fig. \subref*{fig:graph-schedule}.





\subsubsection{\underline{Update Process}}\label{subsubsec:update}

\begin{figure}[!htb]
    \vspace{-0mm}
    \centering
    \hspace{-3mm}
    \subfloat[]{
        \begin{minipage}[c]{0.45\linewidth}
            \label{fig:cube1-a}
            \centering
            \includegraphics[width=\linewidth]{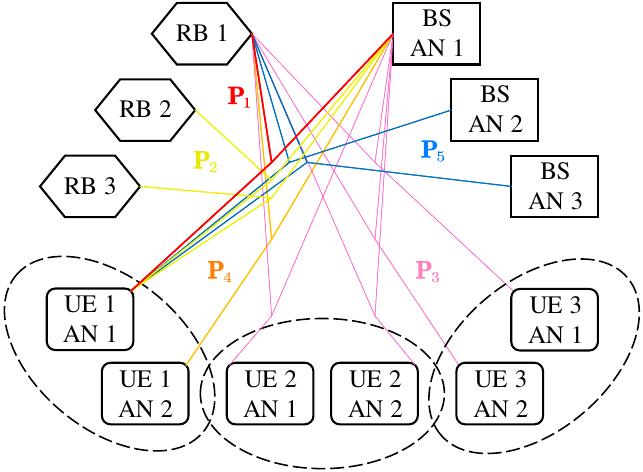}
            \vspace{-4mm}
        \end{minipage}
    }
    \hspace{5mm}
    \subfloat[]{
        \begin{minipage}[c]{0.236\linewidth}
            \label{fig:cube1-b}
            \centering
            \includegraphics[width=\linewidth]{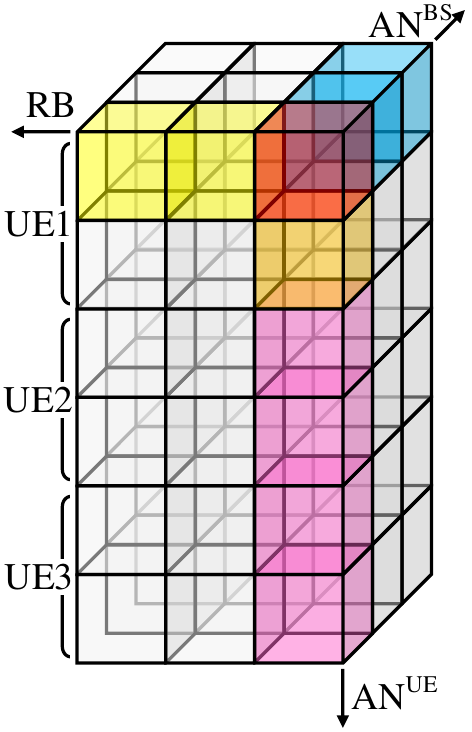}
            \vspace{-7mm}
        \end{minipage}
    }
    \vspace{-0mm}
    \caption{Illustration of the 3D-GNN update process for ${\bf x}_{1,11,1}^l$. $M=3, K=3, N_R=2, N_T=3$. In~(a), ${\bf x}_{1,11,1}^l$ is the hidden representation of the red hyper-edge. ${\bf x}_{1,11,1}^l$ and the representations of four kinds of adjacent hyper-edges are weighted by different trainable weights, indicated by different colors. These weighted representations are summed to update ${\bf x}_{1,11,1}^l$. In~(b), the same update process is illustrated more clearly by using cubes. Each cube denotes the hidden representation of a hyper-edge. The red cube is ${\bf x}^l_{1,11,1}$. All colored cubes are first weighted by the weights of the same colors as in (a) and then summed to update the red one.}
    \label{fig:cube1}
    \vspace{-4mm}
\end{figure}

This process aims to update hidden representations layer by layer. Each representation is updated by aggregating the representations of its adjacent hyper-edges, which are the hyper-edges that differ from the updated one by only one index (i.e., in $m$, $kr$, or $n$). Four kinds of adjacent hyper-edges can be distinguished from the four subscripts. As analyzed in Sec.~\ref{subsec:3d}, linear aggregators enable the 3D-GNN to be generalized well to the sizes of all sets. Therefore, we use four trainable weights to linearly aggregate the representations of adjacent hyper-edges, and use another trainable weight to combine the updated representation itself. As visualized in Fig.~\ref{fig:cube1}, the update process can be expressed~as
\begin{align}
    &{\bf x}^{l+1}_{m,kr,n} \!\!=\! \phi \! \bigg(
    {\bf P}^{l}_1{\bf x}^{l}_{m,kr,n}
    + {\bf P}^{l}_2\frac{1}{M}\sum_{s=1\atop s\neq m}^M{\bf x}^{l}_{s,kr,n}+{\bf P}^{l}_3\frac{1}{KN_R}\cdot  \label{eq:updatex}\\[-2.5mm]
    &\sum_{t=1\atop t\neq k}^K\sum_{u=1}^{N_R}{\bf x}^{l}_{m,tu,n}\!+\!{\bf P}^{l}_4\frac{1}{N_R}\sum_{u=1\atop u\neq r}^{N_R}{\bf x}^{l}_{m,ku,n}
    \!+\! {\bf P}^{l}_5\frac{1}{N_T}\sum_{v=1\atop v\neq n}^{N_T}{\bf x}^{l}_{m,kr,v}\bigg), \nonumber
\end{align}
where ${\bf P}^{l}_j\in\mathbb{R}^{C_{l+1}\times C_l},\ j=1,\cdots,5$ are trainable weights and $\phi(\cdot)$ is an activation function.

In \eqref{eq:updatex}, the same weights ${\bf P}^l_j,j=1,\cdots,5$ are used for all hyper-edges, and their dimensions are independent of the graph size. This parameter-sharing allows the update equation in \eqref{eq:updatex} to be applied to the hidden representation of every hyper-edge in a graph of arbitrary size in the inference phase.


\subsubsection{\underline{Output Process}}

This process aims to produce action $\bf{B}$ from the hidden representations, whose size is $C_{L_S}=1$, in the last layer, namely ${x}^{L_S}_{m,kr,n} \in \mathbb{R}$. The representations are defined on hyper-edges, while the action is defined on edges. In other words, all representations compose a four-dimensional array ${\bf X}^{L_S} \in \mathbb{R}^{M\times KN_R\times N_T\times C_{L_S}}$, but all actions compose ${\bf B}\in \{0,1\}^{M\times K' \times K}$. Hence, we first compress the high-dimensional representation over $N_T$ and $N_R$ as
\begin{align} \label{E:compress}
    {z}_{m,k} = \frac{1}{N_RN_T}\sum_{r=1}^{N_R} \sum_{n=1}^{N_T} {x}^{L_S}_{m,kr,n}\in \mathbb{R}.
\end{align}
With ${z}_{m,k}$, we can obtain the action ${\bf B}$ in both the testing and training phases as follows.


\textbf{Testing phase:} For each RB, the $K'$ basis vectors in ${\bf B}_m$ are determined by the indices corresponding to the largest $K'$ values in ${z}_{m,k}, k=1,\cdots,K$. In particular, the $k$th element in the $k'$th basis vector is given by ${b}_{m,k',k} = {\rm Top}_{k'}(z_{m,k})$, where ${\rm Top}_{k'}({z}_{m,k}) = 1$ if ${z}_{m,k}$ is the $k'$th largest value among the $K$ scalars ${z}_{m,k}, k=1,\cdots,K$, and ${\rm Top}_{k'}({z}_{m,k}) = 0$ otherwise. 

In this way, the constraint in \eqref{prob:1e'} is satisfied since $K'$ different basis vectors are provided on each RB.

\textbf{Training phase:} To enable back-propagation of gradients, $K'$ basis vectors need to be relaxed as continuous vectors. To this end, we adopt the function ${\rm SoftTop}_{k'}(\cdot)$ \cite{top-k}, which yields $K'$ relaxed basis vectors $\tilde{\bf b}_{m,k'} \!\!\in\! (0,1)^{1\times K}$ one by one based on the values of ${z}_{m,k}, k=1,\cdots,K$. 
${\rm SoftTop}_{k'}(\cdot)$ requires $K'$ iterative rounds, each indexed by $k'$. 
Specifically, for $k'=1$, each element in the first relaxed basis vector is given by $\tilde{b}_{m,1,k} = {\rm Softmax}_\tau({z}_{m,k}) = \exp({z}_{m,k}/\tau)/(\sum_{i=1}^K\exp({z}_{m,i}/\tau))$,
where ${\rm Softmax}_\tau(\cdot)$ is a temperature-parameterized function that generates a probability, and $\tau>0$ is the temperature controlling the smoothness of the output distribution. As the training epochs increase, the values of $\tau$ decrease, making the elements in the relaxed basis vector close to discretized values, i.e., with more concentrated distributions. This approach reduces the discrepancy between the outcomes during training and testing~phases. 

Next, for each $k'>1$, the values of ${z}_{m,k}, k=1,\cdots,K$ are adjusted based on the probability obtained in the previous round as ${z}_{m,k}^{(k')} =  {z}_{m,k}^{(k'-1)} + \log(1-\tilde{b}_{m,k'-1,k})$, where ${z}_{m,k}^{(1)} = {z}_{m,k}$. Then, the $k'$th relaxed basis vector is $\tilde{b}_{m,k',k} = {\rm Softmax}_\tau({z}_{m,k}^{(k')})$.
After $K'$ rounds of this process, we can obtain $K'$ relaxed basis vectors on each~RB.



\subsubsection{\underline{A Sequence of GNNs to Enhance Scheduler Module}}
\begin{figure}
	\centering
	\vspace{-0mm}
	\includegraphics[width=0.8\linewidth]{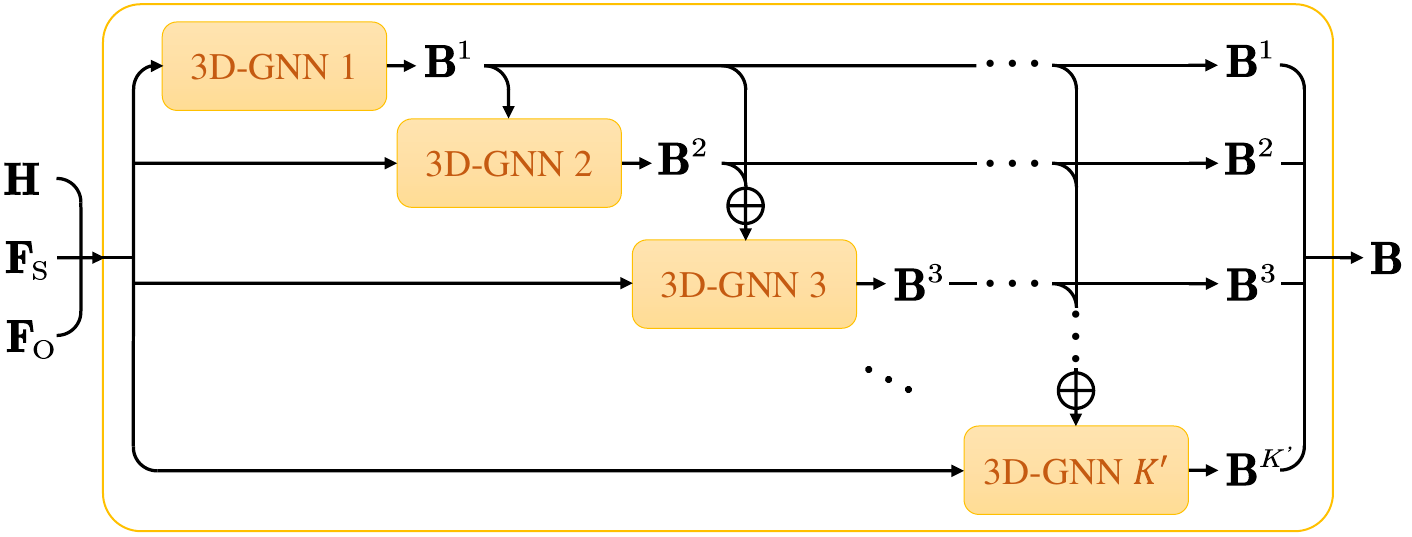}
	\vspace{-2mm}
	\caption{A sequence of GNNs. They can replace the 3D-GNN in the scheduler module of NGNN as an enhancement. After the replacement, the whole architecture is referred to as SGNN.  
    }
        \label{fig:seq}
	\vspace{-3mm}
\end{figure}

As mentioned in Sec.~\ref{subsec:3d}, a single GNN cannot well learn the non-SPSD scheduling policy in the scenario with densely distributed users. To address it, we design a scheduler module consisting of $K'$ 3D-GNNs, as shown in Fig.~\ref{fig:seq}.

The idea is to let each GNN merely choose a single user on every RB, i.e., the $k'$th GNN generates basis vectors ${\bf b}_{m,k'}$, $m=1,\cdots,M$. Meanwhile, each subsequent GNN inputs not only the original features but also all the basis vectors generated by the preceding GNNs.
This design ensures that once one user is scheduled, particularly the one with similar channels to others, the inputs to the subsequent GNNs become different. This method effectively reduces the similarity of inputs for one GNN, thereby avoiding performance degradation for learning non-SPSD policy.

Similar to the GNN designed in Sec.~\ref{subsec:scheModule}, herein each GNN also includes the input, update, and output processes. The update process is similar to that described in Sec.~\ref{subsubsec:update} but with different trainable weights. 
However, the input and output processes require redesign. Specifically, each GNN requires not only the input of channel features but also the scheduling decisions from the preceding GNNs. Moreover, each GNN outputs only one scheduling decision per RB rather than all $K'$ scheduling decisions simultaneously. We next detail the two processes.

\emph{Input process:} Let ${\bf x}^{k',1}_{m,kr,n}\in\mathbb{R}^{C_1}$ denote the first-layer representation of a hyper-edge connecting the $m$th RB, the $kr$th AN$^{\rm UE}$, and the $n$th AN$^{\rm BS}$ vertices for the $k'$th GNN.
For the first GNN, ${\bf x}^{1,1}_{m,kr,n} = [{\rm Re}(({\bf H})_{m,kr,n})$, ${\rm Im}(({\bf H})_{m,kr,n})$, ${F}_{\text{S}{m,k}}$, ${F}_{\text{O}{m,kr}}]^T \in \mathbb{R}^{4}$. For the subsequent GNNs, the inputs further include the summation of outputs from preceding GNNs. Specifically, the input for the $k'$th GNN is ${\bf x}^{k',1}_{m,kr,n} = [{\rm Re}(({\bf H})_{m,kr,n})$, ${\rm Im}(({\bf H})_{m,kr,n})$, $ {F}_{\text{S}{m,k}}$, ${F}_{\text{O}{m,kr}}$, $\sum_{j=1}^{k'-1}{b}_{m,j,k}]^T \in \mathbb{R}^{5}$, $k'=2,\cdots, K'$.

\emph{Output process:} Each GNN generates a basis vector for every RB. For the $k'$th GNN, like in \eqref{E:compress}, the hidden representation in the last layer ${x}^{k',L_S}_{m,kr,n}\in\mathbb{R}$ is first compressed as ${z}_{m,k}^{k'} = \sum_{r=1}^{N_R}\sum_{n=1}^{N_T}$ ${x}^{k',L_S}_{m,kr,n}/N_RN_T\in\mathbb{R}$.
With ${z}_{m,k}^{k'}$, in the \emph{testing phase}, the $k$th element in the basis vector is discretized by ${b}_{m,k',k} = {\rm Onehot}({z}_{m,k}^{k'})$, where ${\rm Onehot}({z}_{m,k}^{k'})=1$ if ${z}_{m,k}^{k'}\geq {z}_{m,i}^{k'}$ for $\forall i\in\{1,\cdots,K\}$, and ${\rm Onehot}({z}_{m,k}^{k'})=0$ otherwise. In Fig. \ref{fig:seq}, the $M$ basis vectors yielded by the $k'$th GNN compose a matrix ${\bf B}^{k'}\in \{0,1\}^{M\times K}$. In the \emph{training phase}, the relaxed basis vector is given by $\tilde{b}_{m,k',k} ={\rm Softmax}_\tau({z}_{m,k}^{k'})$.

By introducing the sequence of GNNs, each employing a discretization step, the scheduler module learns a non-continuous function that produces very different outcomes for users with similar channels. 
After jointly training with the precoder module to maximize SE, the $K'$ GNNs will schedule different users on the same RB, ensuring that the constraint in \eqref{prob:1e'} is satisfied with high probability.

Although SGNN is designed for scenarios with dense user population, it can also be applied in other scenarios, albeit with higher complexity compared to NGNN.


\subsection{Design of Precoder Module}

In this subsection, we design the three processes of the 3D-GNN in the precoder module, similar to the scheduler module in Sec.~\ref{subsec:scheModule}. In the update process, we devise a novel attention mechanism to enhance its generalization performance to the number of~users.

As discussed in Sec.~\ref{subsec:3d}, the precoding policy is SPSD on AN$^{\rm BS}$, AN, and RB sets, but non-SPSD on user set. It indicates that GNNs are not generalizable to the number of users without a judicious design, even if they are permutation equivariant to users (that is, they can be applied to graphs with different number of users \cite{RGNN}).



Inspired by the design in \cite{LSJ_TWC, modelGNN} for precoding in narrow-band MU-MISO systems, the generalization performance to varying $K$ can be improved by weighting the information before aggregation, where the weights can reflect the strength of MUI. We propose an attention coefficient for precoding in wideband MU-MIMO systems. Different from \cite{LSJ_TWC, modelGNN}, the attention coefficient is computed on each RB and needs to reflect the total MUI on multiple receive antennas at each user.

The precoder module is an $L_P$-layer 3D-GNN, where the hidden representation of the hyper-edge connecting the $m$th RB, the $k'r$th AN$^{\rm UE}$, and the $n$th AN$^{\rm BS}$ in the $l$th layer is denoted as ${\bf y}^{l}_{m,k'r,n}\in \mathbb{R}^{D_l}$, $l=1,\cdots,L_P$. There are totally $MK'N_RN_T$ (or $MKN_RN_T$ if $K<K'$) hidden representations in a layer.


We next detail the three processes of the 3D-GNN.

\subsubsection{\underline{Input Process}}



The representations in the first layer are the channels of scheduled users defined on the hyper-edges, as shown in Fig. \subref*{fig:graph-precode}. Each representation is expressed in the real-valued form as ${\bf y}^{1}_{m,k'r,n} = [{\rm Re}(({\bf H}')_{m,k'r,n}), {\rm Im}(({\bf H}')_{m,k'r,n})]^T \in \mathbb{R}^{2}$. 


\subsubsection{\underline{Update Process with Attention Mechanism}}\label{subsubsec:update-atten}
\begin{figure}[!htb]
    \vspace{-4mm}
    \centering
    \hspace{-0mm}
    \subfloat[]{
        \begin{minipage}[c]{0.45\linewidth}
            \label{fig:cube2-a}
            \centering
            \includegraphics[width=\linewidth]{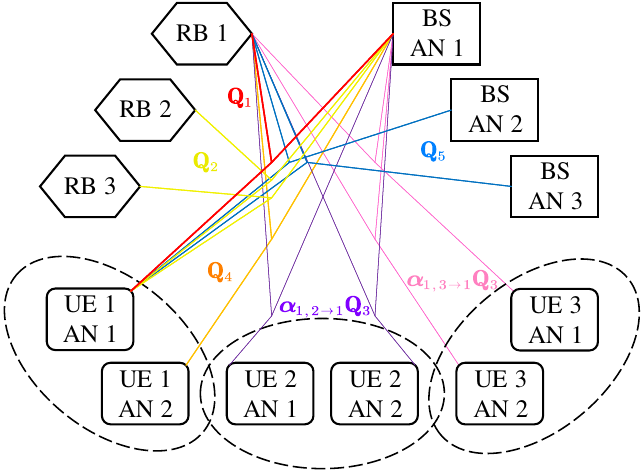}
            \vspace{-4mm}
        \end{minipage}
    }
    \hspace{-5mm}
    \subfloat[]{
        \begin{minipage}[c]{0.5\linewidth}
            \label{fig:cube2-b}
            \centering
            \includegraphics[width=\linewidth]{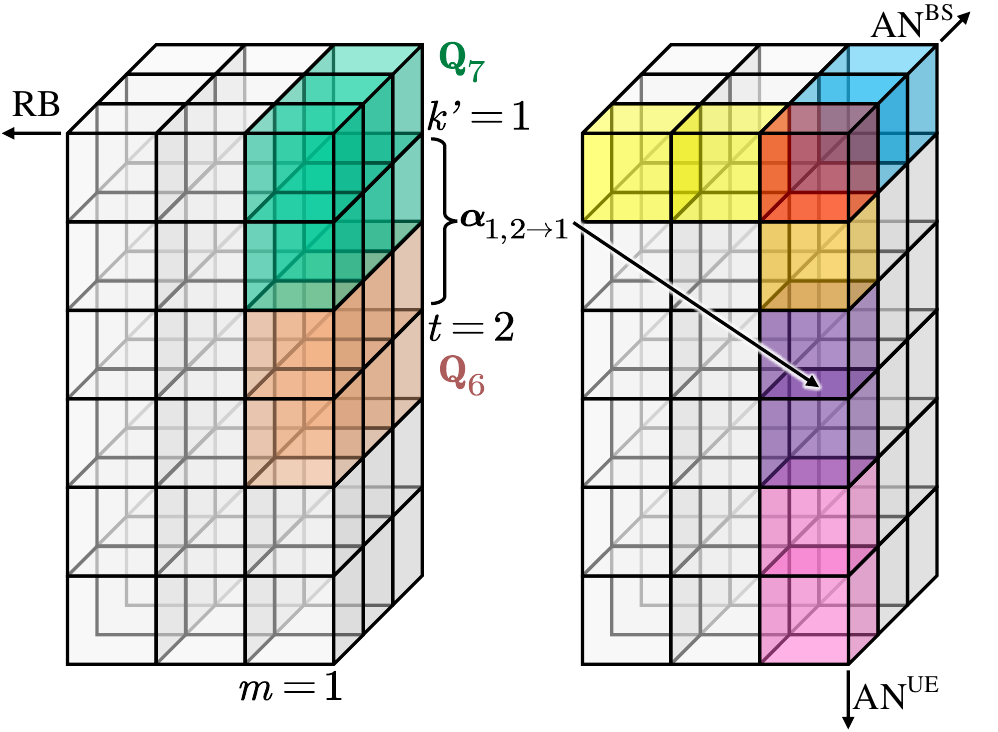}
            \vspace{-7mm}
        \end{minipage}
    }
    \vspace{-2mm}
    \caption{Illustration of the 3D-GNN update process for ${\bf y}_{1,11,1}^l$ with attention mechanism. $M=3, K'=3, N_R=2, N_T=3$. In~(a), ${\bf y}_{1,11,1}^l$ is the hidden representation of the red hyper-edge, which is updated by summing itself and the representations of adjacent hyper-edges with different weights. Only the representations of different users are further weighted by corresponding attention coefficients. In (b), on the left, the computation of attention coefficient ${\bm \alpha}^l_{1,2\rightarrow1}$ is illustrated. On the right, all colored cubes are summed to update the red one (i.e., ${\bf y}^l_{1,11,1}$), where ${\bm \alpha}^l_{1,2\rightarrow1}$ is used to weight the representations of the 2nd user on the 1st RB (the two cubes in purple color).}
    \label{fig:cube2}
    \vspace{-2mm}
\end{figure}

To enable the generalizability to the number of users, $K$, we introduce an attention coefficient ${\bm \alpha}^l_{m,t\rightarrow k'} \in \mathbb{R}^{D_{l+1}}$ to measure the MUI strength from the $t$th user to the $k'$th user on the $m$th RB, when updating the hidden representations of the hyper-edges connecting the antennas at the $k'$th user, where $t\in\{1,\cdots,K'\}\backslash \{k'\}$. 
${\bm \alpha}^l_{m,t\rightarrow k'}$ is then multiplied by the aggregated representations connecting the antennas at the $t$th user.
The other three kinds of representations of adjacent hyper-edges corresponding to the sets of RBs and antennas at both BS and the $k'$th user are aggregated linearly, which enables the generalizability due to the aforementioned SPSD property. As visualized in Fig. \ref{fig:cube2}, the hidden representation of each hyper-edge in the $(l+1)$th layer is updated as
\begin{align}
    {\bf y}^{l+1}_{m,k'r,n} \!\!=\! \phi \! &\bigg(
    {\bf Q}^{l}_1{\bf y}^{l}_{m,k'r,n}
    + {\bf Q}^{l}_2\frac{1}{M}\sum_{s=1\atop s\neq m}^M{\bf y}^{l}_{s,k'r,n}  \nonumber \\[-1.5mm]
    &+\frac{1}{K'}\sum_{t=1\atop t\neq k'}^{K'}{\bm \alpha}^l_{m,t\rightarrow k'} \odot {\bf Q}^{l}_3\frac{1}{N_R}\sum_{u=1}^{N_R}{\bf y}^{l}_{m,tu,n}\label{eq:updatey}\\[-1.5mm]
    &+{\bf Q}^{l}_4\frac{1}{N_R}\sum_{u=1\atop u\neq r}^{N_R}{\bf y}^{l}_{m,k'u,n}
    + {\bf Q}^{l}_5\frac{1}{N_T}\sum_{v=1\atop v\neq n}^{N_T}{\bf y}^{l}_{m,k'r,v}\bigg), \nonumber
\end{align}
where the attention coefficient is obtained as 
\begin{align}\label{eq:alpha}
    {\bm \alpha}^l_{m,t\rightarrow k'} = \tanh\bigg(\frac{1}{N_T} \sum_{v=1}^{N_T} \Big({\bf Q}^{l}_6\frac{1}{N_R}\sum_{u=1}^{N_R}{\bf y}^{l}_{m,tu,v} \nonumber \\[-1.5mm]
    \odot  {\bf Q}^{l}_7\frac{1}{N_R}\sum_{u=1}^{N_R}{\bf y}^{l}_{m,k'u,v} \Big)\bigg),
\end{align}
${\bf Q}^{l}_j\in \mathbb{R}^{D_{l+1}\times D_l}, \ j=1,\cdots,7$ are trainable weights, and $\tanh(\cdot)$ is introduced to restrict the value of attention coefficient for assisting the generalization to the number of antennas.

If ${\bf Q}^{1}_6={\bf Q}^{1}_7={\bf I}$ and $N_R=1$ in \eqref{eq:alpha}, the outer-layer summation over $N_T$ in the first layer is the inner product between the channels of the $t$th and $k'$th users, whose value increases with the correlation between the channels. Since each user receives a single data stream, the attention coefficient between the averaged representations of the hyper-edges connecting two users over their receive antennas can be regarded as the total interference between them on an RB. With the aid of such attention coefficients, the importance of the information from different users on the hidden representation of a certain user can be well distinguished.

The weights ${\bf Q}^l_j,\ j=1,\cdots,7$ in \eqref{eq:updatey} and \eqref{eq:alpha} are independent of the size of the precoding graph, and are shared across (i.e., the same for) all hyper-edges. This allows the update equation in \eqref{eq:updatey} to be applicable to the hidden representation of every hyper-edge in the precoding graphs with varying sizes during inference.

\subsubsection{\underline{Output Process}}
The actions of the analog combiner and hybrid precoders are obtained from the hidden representations in the last layer after being normalized to satisfy constraints. The number of elements in each ${\bf y}^{L_P}_{m,k'r,n}$ is set to $D_{L_P}=4N_{\text{RF}}+2$.



The real and imaginary parts of $\tilde{\bf W}_\text{RF}$ come from the first and second $N_{\rm RF}$ elements, which are respectively
\begin{align}
  {\rm Re}((\tilde{\bf W}_\text{RF})_{n,j}) &\!=\! \frac{1}{MKN_R}\!\sum_{m=1}^M\!\sum_{k'=1}^{K}\!\sum_{r=1}^{N_R} ({\bf y}^{L_P}_{m,k'r,n})_j\in\mathbb{R}
\end{align}
and 
\begin{align}
  {\rm Im}((\tilde{\bf W}_\text{RF})_{n,j}) &\!=\! \frac{1}{MKN_R}\!\sum_{m=1}^M\!\sum_{k'=1}^{K}\!\sum_{r=1}^{N_R} ({\bf y}^{L_P}_{m,k'r,n})_{N_{\rm RF}+j}\in\mathbb{R},
\end{align}
for $j = 1,\cdots, N_{\text{RF}}$. Then, to satisfy constraint \eqref{prob:1b}, $({\bf W}_\text{RF})_{n,j} = (\tilde{\bf W}_\text{RF})_{n,j}/|(\tilde{\bf W}_\text{RF})_{n,j}|$.

The real and imaginary parts of $\tilde{\bf W}'_{\text{BB}}$ come from the third and fourth $N_{\rm RF}$ elements, which are respectively
\begin{align}
  {\rm Re}((\tilde{\bf W}'_{\rm BB})_{m,j,k'}) &\!=\! \frac{1}{N_RN_T} \!\sum_{r=1}^{N_R}\!\sum_{n=1}^{N_T}({\bf y}^{L_P}_{m,k'r,n})_{2N_{\rm RF}+j}\!\in\!\mathbb{R} \end{align}
and
\begin{align}
  {\rm Im}((\tilde{\bf W}'_{\rm BB})_{m,j,k'}) &\!=\! \frac{1}{N_RN_T} \!\sum_{r=1}^{N_R}\!\sum_{n=1}^{N_T}({\bf y}^{L_P}_{m,k'r,n})_{3N_{\rm RF}+j}\!\in\!\mathbb{R},
\end{align}
for $j = 1,\cdots, N_{\text{RF}}$. Then, to satisfy power constraint \eqref{prob:pa}, ${\bf w}'_{{\rm BB}m,k'} \!=\! \tilde{\bf w}'_{{\rm BB}m,k'}\!\! \sqrt{P_{tot}/ \sum_{m=1}^M\sum_{k'=1}^{K'}||{\bf W}_{\rm RF}\tilde{\bf w}'_{{\rm BB}m,k'}||_2^2}$.

The real and imaginary parts of $\tilde{\bf v}'_{\text{RF}}$ come from the last two elements, which are respectively
\begin{align}
  {\rm Re}((\tilde{\bf v}'_{\rm RF})_{k'r}) &\!=\! \frac{1}{MN_T}\!\sum_{m=1}^M\!\sum_{n=1}^{N_T} ({\bf y}^{L_P}_{m,k'r,n})_{4N_{\rm RF}+1}\in \mathbb{R}
\end{align}
and 
\begin{align}
  {\rm Im}((\tilde{\bf v}'_{\rm RF})_{k'r}) &\!=\! \frac{1}{MN_T}\!\sum_{m=1}^M\!\sum_{n=1}^{N_T} ({\bf y}^{L_P}_{m,k'r,n})_{4N_{\rm RF}+2}\in \mathbb{R}.
\end{align}
Then, to satisfy constraint \eqref{prob:pc}, $({\bf v}'_{\rm RF})_{k'r} \!=\! (\tilde{\bf v}'_{\rm RF})_{k'r}/|(\tilde{\bf v}'_{\rm RF})_{k'r}|$.

\noindent\textbf{Remark 1:} The proposed architecture can be extended to the setup where each user receives $N_S>1$ data streams, and both data streams and users need to be selected \cite{LISA}. In this case, the scheduling matrix becomes ${\bf A}\in \{0,1\}^{M\times KN_S}$ and its version with basis vectors becomes ${\bf B}\in \{0,1\}^{M\times N_D\times KN_S}$, where $N_D$ is the total number of data streams. 
The number of scheduled users $K'$ ranges from $\lceil N_D/N_S\rceil$ to $N_D$, where $\lceil \cdot \rceil$ stands for the ceiling operation.
For the GNNs in the two modules, the weight-sharing and attention mechanism should be designed by taking the permutation of data streams into consideration \cite{LSJ_TWC}, where the design of attention mechanism remains an open problem.

\subsection{Training Phase and Test Phase}
Next, we show how to train the architecture (including NGNN and SGNN) and apply it for inference.

\subsubsection{Training Phase}
The precoder module should be adaptable to different channel distributions of the scheduled users, meanwhile the performance of a scheduler depends on the precoder. Moreover, an adequate precoder is a prerequisite for training the scheduler. If a random precoder is used, then the scheduler cannot find the users to maximize the SE. Hence, the precoder module should be pre-trained first, and then fine-tuned by jointly training the two modules.

\paragraph{Training the Precoder Module} To allow the GNN in the precoder module to fast adapt to channel distribution determined by the scheduler module, the GNN is pre-trained. Each sample for the pre-training comprises the channels of $K'$ users selected from $K$ users by a simple scheduling method, i.e., choosing $K'$ users with the strongest channels on each RB.

The loss function for one sample is $\mathcal{L}=-\frac{1}{M} \sum_{m=1}^M \sum_{k'=1}^{K'} R'_{m,k'}$. Then, the gradient of the averaged loss over a batch of samples is back-propagated to update the trainable weights in the GNN.

\paragraph{Training the Scheduler Module} The scheduler module is trained by the samples each with the channels of $K$ candidate users. These samples go through the entire architecture, composed of the trainable scheduler and the pre-trained precoder with frozen weights. The loss function is also $\mathcal{L}$, but the scheduled users may change during training. The gradient of the averaged loss over a batch of samples is back-propagated to update the weights in the GNN (or GNNs) in the scheduler module. 


\paragraph{Jointly Training the Two Modules} The two modules are trained jointly by the same samples as those for training the scheduler module. These samples go through the entire architecture, where the weights of all GNNs in both modules have been pre-trained. The loss function remains $\mathcal{L}$. The gradient of the averaged loss over a batch of samples is back-propagated to fine-tune the weights in all GNNs.

\subsubsection{Test Phase} If $K>K'$, the inference process is shown in Fig. \ref{fig:arch}. The scheduler module yields scheduling outcome, and the precoder module produces the corresponding combiner and precoders. If $K\leq K'$, the scheduler module is inactive, and the combiner and precoders can be obtained from the precoder module with ${\bf H}' = {\bf H}$.


The GNNs in the scheduler and precoder modules can be applied to scheduling and precoding graphs of varying sizes, as discussed in Sec.~\ref{subsubsec:update} and \ref{subsubsec:update-atten}. To be more precise, the well-trained GNNs perform well when learning the scheduling and precoding policies with different numbers of RBs, candidate users, and antennas at the BS and each user during the test phase.



\subsection{Computational Complexity Analysis}
The computational complexity for inference is measured in floating point operations (FLOPs).

\subsubsection{3D-GNN in Scheduler Module}
The 3D-GNN in NGNN and each one in SGNN have the same expression of complexity but with different hyper-parameters.
We analyze the update process in Sec.~\ref{subsubsec:update}, because the complexities of input and output processes are ignorable. In the first term on the right-hand side of \eqref{eq:updatex}, multiplying the matrix by a vector requires $C_{l+1}C_l$ multiplications and $C_{l+1}(C_l-1)$ additions. This term needs to be computed for all the $MKN_RN_T$ hyper-edges. The second term can be rewritten as $\tilde{\bf P}^{l}_2\frac{1}{M}\sum_{s=1}^M{\bf x}^{l}_{s,kr,n}$ such that it can be reused across RBs, where $C_{l+1}(C_l+1)$ multiplications and $C_{l+1}(C_l-1)+C_l(M-1)$ additions are required. This term needs to be computed $KN_RN_T$ times for all hyper-edges. The FLOPs for the other three terms can be derived similarly, and the sum of five terms involves $4C_{l+1}$ additions. Thereby, the total FLOPs required by one layer of a 3D-GNN in the scheduler module is $C_{l+1}(2C_l-1)MKN_RN_T + 2C_{l+1}C_l(KN_RN_T+MN_T+MKN_T+MKN_R)+ C_l[(M-1)KN_RN_T+(KN_R-1)MN_T+(N_R-1)MKN_T+(N_T-1)MKN_R] + 4C_{l+1}$.
The overall complexity of SGNN is higher than that of NGNN because SGNN employs $K'$ 3D-GNNs in the scheduler module.

\subsubsection{3D-GNN in Precoder Module}
The 3D-GNN is same for both the NGNN and SGNN.
Except the third term in \eqref{eq:updatey}, the update process is analogous to \eqref{eq:updatex}. The attention coefficient in \eqref{eq:alpha} is realized by two matrix-vector multiplications and a matrix-matrix (with sizes of $K'\times N_T$ and $N_T\times K'$) multiplication. Then, to incorporate attention coefficient, the third term needs a matrix-vector multiplication and a matrix-matrix (with sizes of $K'\times K'$ and $K'\times N_T$) multiplication. Thereby, the total FLOPs required by one layer of the 3D-GNN in the precoder module is $D_{l+1}(2D_l-1)MK'N_RN_T + 2D_{l+1}D_l(K'N_RN_T+4MK'N_T+MK'N_R) + D_l[(M-1)K'N_RN_T+(N_R-1)MK'N_T+(N_T-1)MK'N_R] + D_{l+1}MN_T(4K'^2+4K'N_R-K'+1) $. 

The order of magnitude of FLOPs for the scheduler and precoder modules in the proposed architecture, including both the NGNN and SGNN, are provided in Table \ref{table:FLOPs}, where $C_{l+1}=C_l=C$ and $D_{l+1}=D_l=D$ are assumed for notational simplicity.
For comparison, we also list the results of several numerical algorithms for wideband scheduling, wideband hybrid precoding, or jointly designing the two, where $C_{pre}$ represents the complexity of the used precoding algorithm.
It is evident that both modules in the proposed architecture are with the lowest complexity (except the ``Strongest'' scheduling) due to the lack of high-order terms of $N_T$ and $N_R$.

\begin{table}[htb!]
\setlength\tabcolsep{0.8pt}
\centering
\vspace{-0mm}
\caption{Computational Complexity}\label{table:FLOPs}
\vspace{-2mm}
\footnotesize
\begin{threeparttable}
    \begin{tabular}{c|c}
        \hline\hline
        \bf Scheduling Methods & \bf Asymptotic Complexity\\ \hline
        Scheduler in NGNN & $O(MKN_RN_TL_SC^2)$ \\ \hline
        Scheduler in SGNN & $O(MKK'N_RN_TL_SC^2)$ \\ \hline
        Multicarrier SUS \cite{MSUS} & $O(MKK'N_R^2N_T^2 )$ \\ \hline
        Strongest & $O(MK(N_RN_T+\log K'))$ \\ \hline\hline
        \bf Precoding Methods & \bf Asymptotic Complexity\\ \hline
        Precoder in NGNN/SGNN & $O(MK'N_RN_TL_PD(K'+D))$ \\ \hline
        SDR & $O(MK'^2N_{\text{RF}} + MK'N_{\text{RF}}^2+N_T^{4.5})$ \\ \hline
        MO \cite{MOcomplexity} & $O(MN_T^2N_{\text{RF}}+MN_TN_{\text{RF}}^2+MN_{\text{RF}}^3)$ \\ \hline
        OMP \cite{SVDYH} & $O(MK'^2N_{\text{RF}}N_T^2+K'N_R^2)$ \\ \hline
        SVD \cite{SVDYH} & $O(MK'^3N_{\text{RF}}N_T+K'N_R^2)$ \\ \hline
        EIG \cite{EIGLY} & $O(MK'^3+MK'N_{\text{RF}}N_T + N_{\text{RF}}N_T^2)$ \\
        \hline\hline
        \bf Joint Methods & \bf Asymptotic Complexity\\ \hline
        Exhaustive & $O(\binom{K}{K'}^M C_{pre})$ \\ \hline
        Greedy & $O(MKK'C_{pre})$  \\ \hline
        LISA \cite{LISA} & \makecell{ $O(\!MKN_R^2N_T\!\!+\!\!K\min(M^2N_T,\!MN_T^2)\!\!+\!\!r^3\!)$\\  rank of channels $r\leq\min(MN_R,N_T)$} \\
        \hline\hline
    \end{tabular}
\begin{tablenotes}
 \footnotesize
\item Multicarrier SUS: multicarrier semi-orthogonal user selection \cite{MSUS}, SDR: semidefinite relaxation \cite{KLX}, MO: manifold optimization \cite{MO}, OMP: orthogonal matching pursuit \cite{OMP}, SVD: singular value decomposition \cite{SVDYH}, EIG: eigen-decomposition \cite{EIGLY}, LISA: linear successive allocation \cite{LISA}.
\end{tablenotes}
\end{threeparttable}
\vspace{-0mm}
\end{table}

Besides,
if the channel of each user is regarded as a token or a feature of a user vertex, we can show that the dominant term in the number of FLOPs required by Transformer \cite{transformer} and GAT \cite{GAT} is $O(K'^2)$. This indicates that their computational complexities  are comparable to the GNN in the precoder module. The dominant term will be $O(K^2)$ if the joint policy $f_J(\cdot)$ is learned, since the features and actions are considered for all candidate users including the unscheduled ones.

\section{Simulation Results}\label{sec:sim}
In this section, we evaluate the learning performance, size generalizability, inference and training complexities of the two variants of the proposed architecture: NGNN and SGNN.

Consider a non-line-of-sight channel model in urban macro (UMa) scenario\cite{3gpp38901}.
The system and channel parameters are listed in Table \ref{table:scenario}. Users are randomly located in a sector of a cell. Both BS and users are equipped with uniform planar array. Each RB consists of 12 subcarriers and 14 OFDM symbols in a time slot. The noise power on each RB is $\sigma^2 = P_n/M_{max}$, where the noise power in the entire bandwidth $P_n {\rm (dBm)}= N_0 + 10\log_{10}(BW) + N_F$. The pathloss model is $PL{\rm (dB)} = 13.54+39.08\log_{10}(d_{3D})+20\log_{10}(f_c{\rm (GHz)})-0.6(h_U-1.5)$ \cite{3gpp38901}, where $d_{3D}$ is the distance between user and BS. The maximal number of scheduled users is $K'=N_{\text{RF}}$.

\begin{table}[htb!]
\centering
\vspace{-0mm}
\caption{Simulation Setup}\label{table:scenario}
\vspace{-2mm}
\footnotesize
    \begin{tabular}{c|c|c}
        \hline\hline
        \bf Description & \bf Notation & \bf Value\\ \hline
        Carrier frequency & $f_c$ & 28 GHz\\ \hline
        Bandwidth & $BW$ & 400 MHz\\ \hline
        Subcarrier spacing & - & 120 kHz\\ \hline
        Maximal number of RBs & $M_{max}$ & 264\\ \hline
        Default transmit power & $P_{tot}$ & 46 dBm\\ \hline
        Noise spectral density & $N_0$ & -174 dBm/Hz\\ \hline
        Noise figure & $N_F$ & 7 dB \\ \hline
        Number of candidate users & $K$ & 1 to 60\\ \hline
        Number of antennas at the BS & $N_T$ & 8 to 128\\ \hline
        Number of RF chains at the BS & $N_{\text{RF}}$ & 2 to 12 \\ \hline
        Height of BS antennas & - & 25 m\\ \hline
        Number of antennas at each user & $N_R$ & 1 to 8\\ \hline
        Height of user antennas & $h_U$ & 1.5 to 2.5 m\\ \hline
        Cell radius & - & 250 m\\ \hline
        Minimum distance between BS and user & - & 35 m\\ \hline
        User velocity & - & 3 km/h\\ \hline
        Standard deviation of shadowing & - & 6 dB \\
        \hline\hline
    \end{tabular}
    \vspace{-0mm}
\end{table}

We generate 200,000 samples for training and 2,000 samples for testing the DNNs. Activation function is ${\rm ReLU}(\cdot)$, optimizer is Adam, and batch normalization is used. The tuned hyper-parameters are as follows. The batch-size is 50. Initial learning rates are 0.001 and 0.0003 for the precoder and scheduler modules, respectively. The numbers of elements in hidden representations are [4, 64, 64, 64, 64, 1] for the six layers in the GNN for scheduling in NGNN, [5, 32, 32, 1] for the four layers in each GNN in the scheduler module of SGNN, and [2, 128, 128, 128, 128, 128, 128, 2+4$N_{\text{RF}}$] for the eight layers in the GNN for precoding in both variants. The numbers of epochs for training the precoder module, scheduler module, and jointly training the two modules are $E_P=90$, $E_S=10$, and $E_J=100$, respectively. The temperature parameter decays according to $\tau = 0.1+0.4 \exp(-0.02 \times {\rm epoch})$.


\subsection{Learning Performance}
We first evaluate the SE achieved by NGNN and SGNN, by comparing with three numerical algorithms for wideband scheduling and hybrid precoding: LISA \cite{LISA}, Greedy \cite{Bogale}, and SDR \cite{KLX}. ``LISA'' and ``Greedy'' jointly design scheduling and precoding, where ZF constraint is imposed on precoders in ``LISA'' and ZF precoder is employed when selecting users in ``Greedy''. ``SDR'' only optimizes precoding, where the users with strongest channels are scheduled. Since few studies jointly design scheduling and precoding, we further simulate four precoding-only algorithms: SVD \cite{SVDYH}, EIG \cite{EIGLY}, MO \cite{MO}, and OMP \cite{OMP}, which are used together with a multicarrier SUS \cite{MSUS} scheduling algorithm. 

\begin{figure}[!htb]
    \vspace{-0mm}
    \centering
    \subfloat[$N_T=16, N_R=1$.]{
        \begin{minipage}[c]{0.73\linewidth}
            \label{fig:SNR1}
            \centering
            \includegraphics[width=\linewidth]{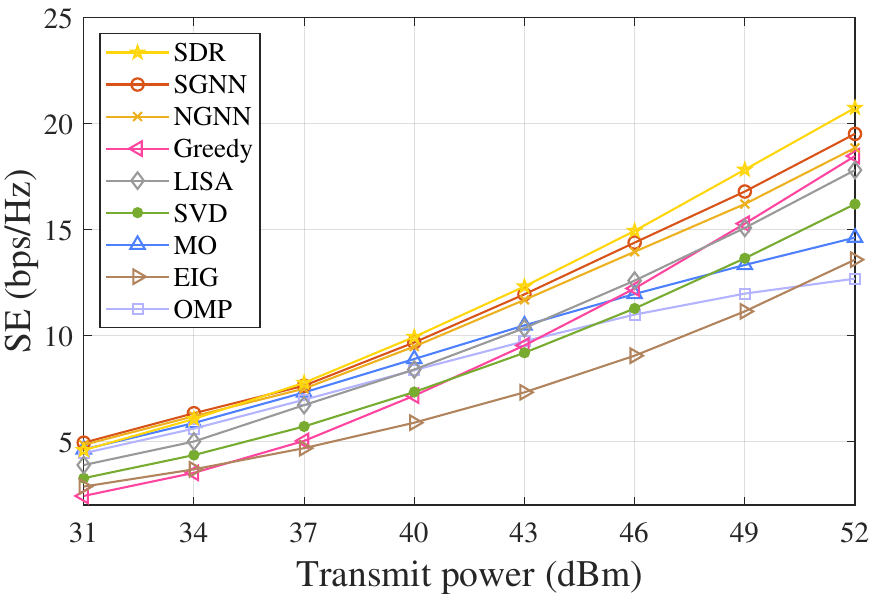}
            \vspace{-5mm}
        \end{minipage}
    }
    \vspace{-3mm}
    \subfloat[$N_T=32, N_R=2$.]{
        \begin{minipage}[c]{\linewidth}
            \label{fig:SNR2}
            \centering
            \includegraphics[width=0.73\linewidth]{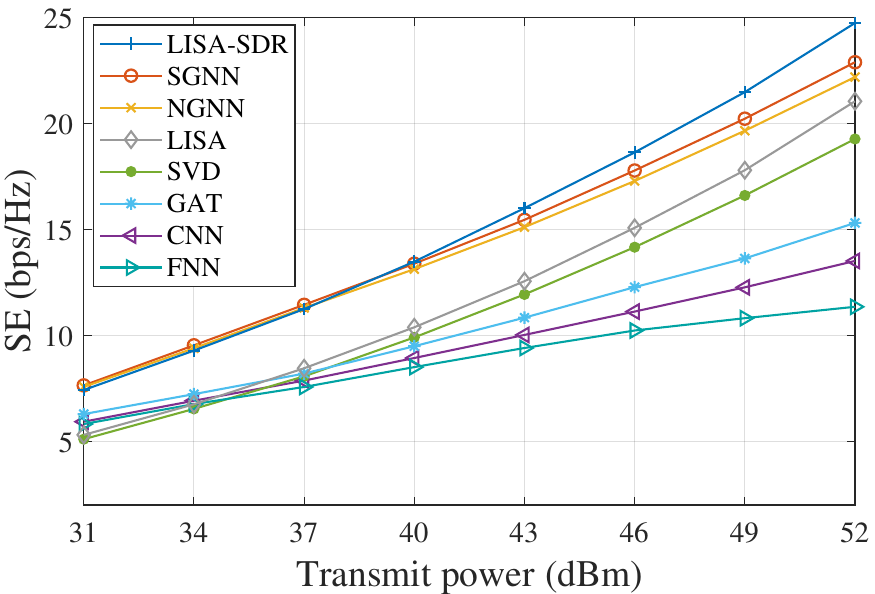}
            \vspace{-1mm}
        \end{minipage}
    }
    \vspace{-0mm}
    \caption{Impact of $P_{tot}$, $M=16, K=10, N_{\text{RF}}=4$.}
    \label{fig:SNR}
    \vspace{-0mm}
\end{figure}

In Fig. \ref{fig:SNR}, we show the impact of signal-to-noise ratio (SNR).
In Fig. \subref*{fig:SNR1}, each user is with a single antenna. It is shown that NGNN and SGNN perform closely to ``SDR'' and outperform other methods. The gap between NGNN and SGNN is minor, because the impact of the non-SPSD property of the scheduling policy is negligible in the considered UMa~scenario.

In the sequel, ``SDR'', ``Greedy'', and ``EIG'' are no longer simulated, since they are designed for single-antenna users. ``MO'' and ``OMP'' are also not simulated anymore, due to their time-consuming iterations and inferior performance. To obtain a benchmark for comparison, we develop an integrated method ``LISA-SDR'', whose scheduling algorithm and analog combiner come from ``LISA'' and precoders come from ``SDR'' based on the equivalent channel, i.e., the channels multiplied by the analog combiner.
Besides, the following three learning-based methods are compared:
\begin{itemize}
\item GAT: All 3D-GNNs in the SGNN are replaced by GATs, which learn over graphs with only user vertices \cite{GATbeijiao}.
\item CNN: Two CNNs, each composed of convolutional layers with kernel size of $3\times 3$ and a fully connected layer \cite{CNNprecode}, serve as the scheduler and precoder modules.
\item FNN: Two FNNs serve as the scheduler and precoder modules.
\end{itemize}

In Fig. \subref*{fig:SNR2}, each user is with two antennas. The superior performance of NGNN and SGNN is evident from the enlarged gaps with other learning methods as SNR increases. 
GAT performs poor owing to its failure to distinguish MUI with properly designed attention mechanism and to leverage the permutation priors of antennas and RBs.
CNN and FNN cannot harness any permutation prior and hence perform worse.

\begin{figure}[htb!]
	\centering
	\vspace{-0mm}
	\includegraphics[width=0.73\linewidth]{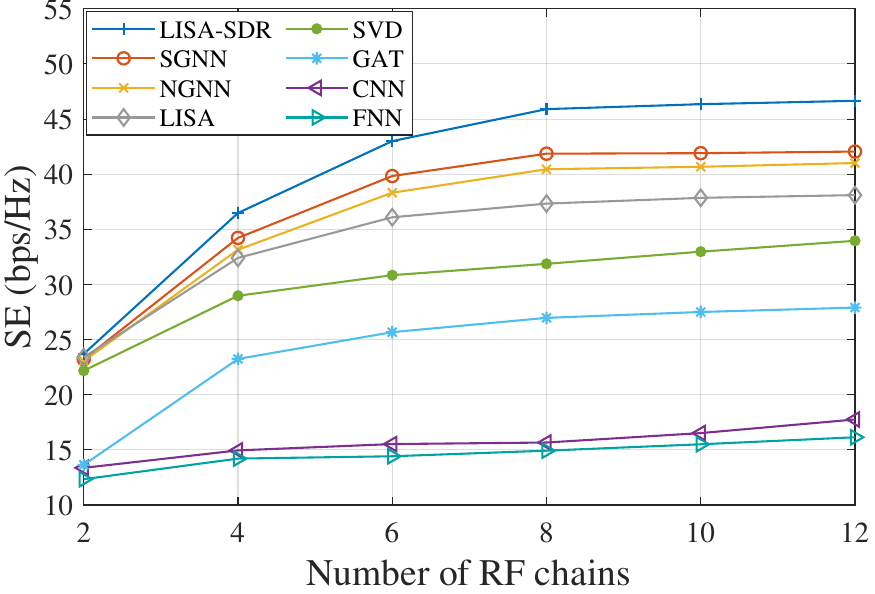}
	\vspace{-3mm}
	\caption{Impact of $N_{\text{RF}}$, $M=4, K=30, N_T=32, N_R=2$.}
       \label{fig:NRF}
	\vspace{-2mm}
\end{figure}

In Fig. \ref{fig:NRF}, we show the impact of the number of RF chains when $P_{tot}=46$ dBm. The result is similar to that in Fig. \subref*{fig:SNR2}.

\begin{figure}[htb!]
	\centering
	\vspace{-0mm}
	\includegraphics[width=0.73\linewidth]{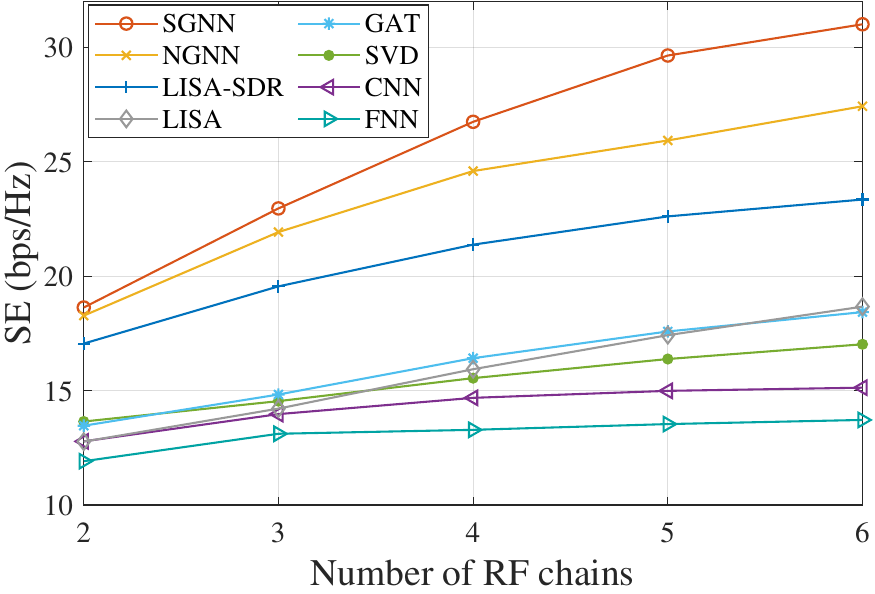}
	\vspace{-3mm}
\caption{Impact of scenario, $M = 16, K = 30, N_T=16, N_R=2$.}
       \label{fig:NRF2}
	\vspace{-2mm}
\end{figure}

In Fig. \ref{fig:NRF2}, we show the gain of the sequential design in SGNN. To this end, we consider a scenario with crowded candidate users, which are located in a 10$\times$10~m$^2$ area at 100~m away from the BS. The channels are generated according to the clustered delay line-A  model\cite{3gpp38901}. 
It shows that SGNN outperforms NGNN more evidently as $N_{\text{RF}}$ increases, since NGNN cannot properly select users with similar channels. Besides, both SGNN and NGNN achieve  higher SE than other~methods.

\subsection{Generalizability to Unseen Problem Scales}
Next, we assess the generalizability of well-trained DNNs, by the ratio of the SE achieved by the learned policies to the SE achieved by ``LISA-SDR'', which can serve as a performance upper bound in the UMa scenario. We also provide the results of ``LISA'' \cite{LISA} and ``SVD''  \cite{SVDYH}, which are simulated for every value of $M$, $K$, $N_T$, or $N_R$.

According to previous analyses, NGNN and SGNN can be generalized to the numbers of RBs, users, AN$^{\rm BS}$s, and AN$^{\rm UE}$s. GAT is generalizable to the number of users, but CNN and FNN are not generalizable to any problem scale. To show the impacts of the model-based features and attention mechanism, two modified SGNNs are also simulated: ``SGNN w/o F'' (that is an SGNN without the model-based features in the scheduler module), and ``SGNN w/o A'' (that is an SGNN without the attention coefficient in the precoder module).

\begin{figure}[htb!]
	\centering
	\vspace{-0mm}
	\includegraphics[width=0.73\linewidth]{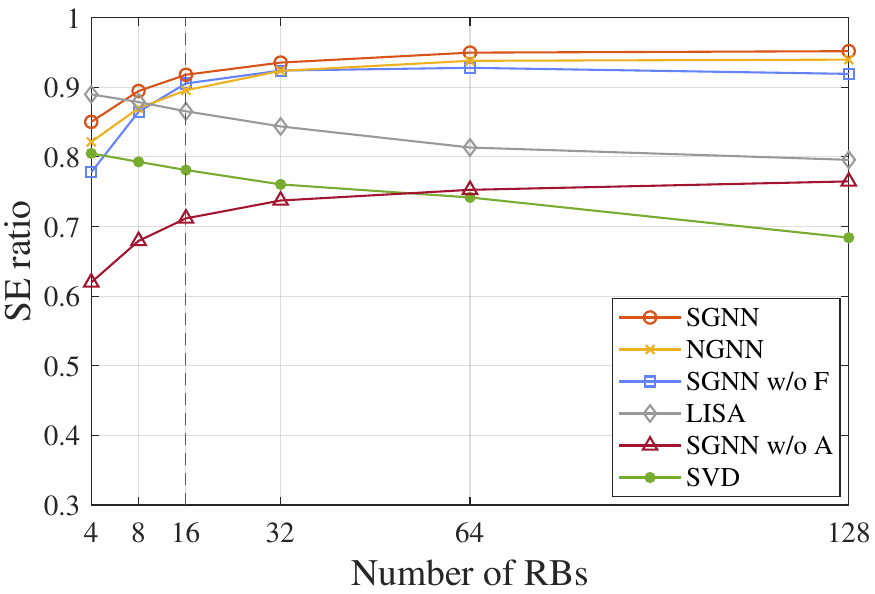}
	\vspace{-3mm}
	\caption{Generalizability to $M$, $K=20, N_{\text{RF}}=4, N_T=16, N_R=2$.}
        \label{fig:genM}
	\vspace{-2mm}
\end{figure}

In Fig. \ref{fig:genM}, we provide the SE ratios achieved by the GNNs that are trained with samples of 16 RBs and tested with samples of 4 to 128 RBs without retraining. As expected, all GNNs can be well generalized to the number of RBs.
The performance of ``LISA'' and ``SVD'' degrades with the enlarged bandwidth, because they involve the matrix factorization of the summation or concatenation of channels across different RBs.

\begin{figure}[htb!]
	\centering
	\vspace{-0mm}
	\includegraphics[width=0.73\linewidth]{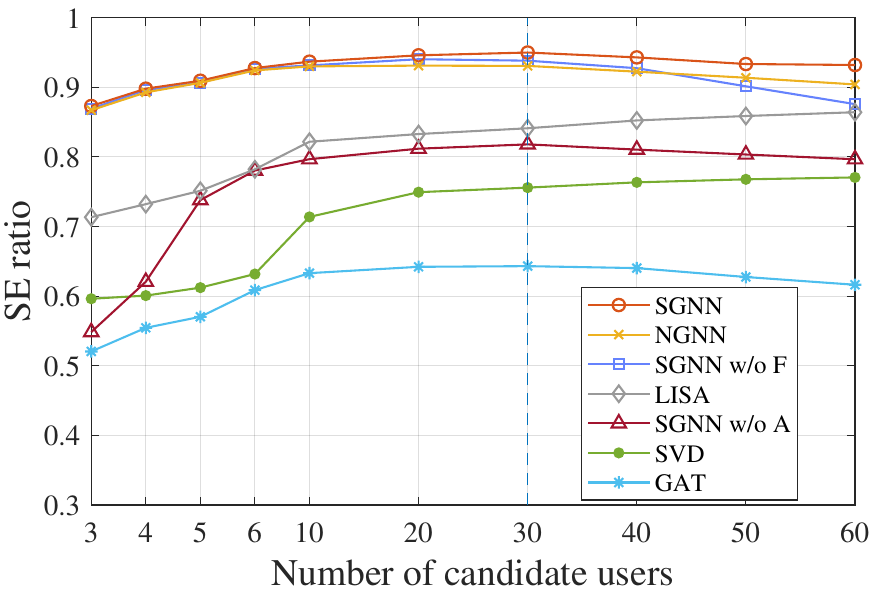}
	\vspace{-3mm}
	\caption{Generalizability to $K$, $M=16, N_{\text{RF}}=6, N_T=16, N_R=2$.}
        \label{fig:genK}
	\vspace{-2mm}
\end{figure}

In Fig. \ref{fig:genK}, we provide the SE ratios achieved by the GNNs that are trained with samples of 30 candidate users, and tested with samples of 3 to 60 users. When $K\leq N_{\text{RF}}$, the generalizability of the precoder module is examined since the scheduler module is inactive. We can see that the proposed attention mechanism significantly improves generalization performance. When $K> N_{\text{RF}}$, the generalizability of the scheduler module is evaluated since precoder always serves $N_{\text{RF}}$ users. We can see that ``SGNN w/o F'' performs worse than SGNN when $K$ is large, 
which indicates the benefit of model-based features for size generalization of scheduling. 

\begin{figure}[htb!]
	\centering
	\vspace{-0mm}
	\includegraphics[width=0.73\linewidth]{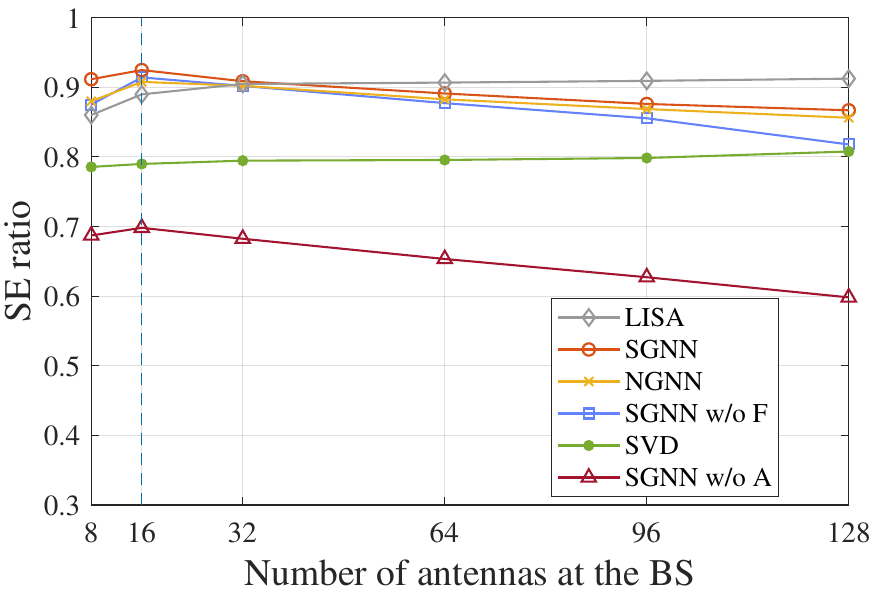}
	\vspace{-3mm}
	\caption{Generalizability to $N_T$, $M=4, K=20, N_{\text{RF}}=4, N_R=2$.}
        \label{fig:genNT}
	\vspace{-2mm}
\end{figure}

\begin{figure}[htb!]
	\centering
	\vspace{-0mm}
	\includegraphics[width=0.73\linewidth]{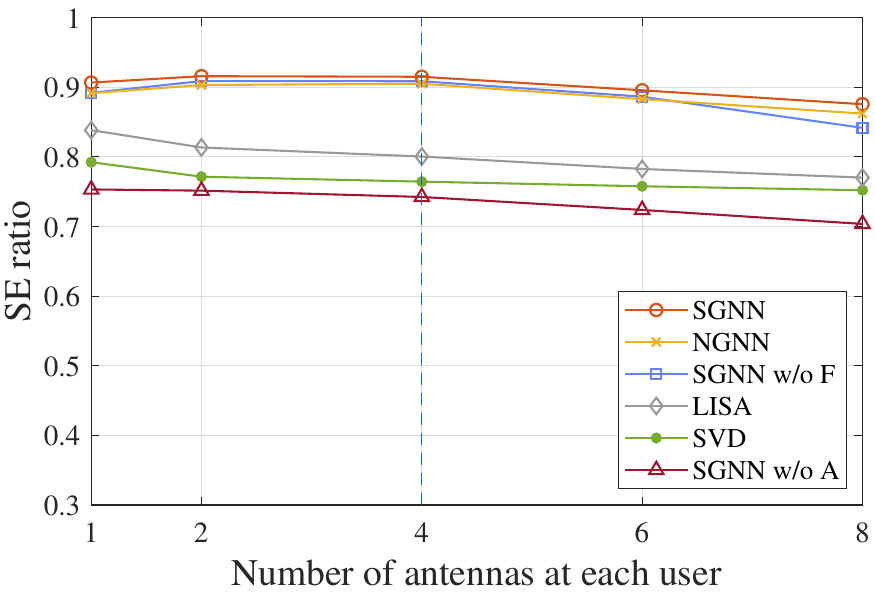}
	\vspace{-3mm}
	\caption{Generalizability to $N_R$, $M=16, K=20, N_{\text{RF}}=4, N_T=16$.}
        \label{fig:genNR}
	\vspace{-2mm}
\end{figure}

In Fig. \ref{fig:genNT}, we provide the SE ratios achieved by the GNNs that are trained with samples of 16 antennas at the BS and tested with samples of 8 to 128 antennas. In Fig. \ref{fig:genNR}, the GNNs are trained with $N_R=4$ and tested with one to eight antennas. In both figures, all GNNs can be well generalized, except ``SGNN w/o F'', which has a slight decline attributed to the changing channel distribution with $N_T$ and $N_R$.

It is noteworthy that the favorable generalization performance to the numbers of RBs, candidate users (for scheduling), AN$^{\rm BS}$s, and AN$^{\rm UE}$s is achieved only with linear aggregators, as shown in the update equations in \eqref{eq:updatex} and \eqref{eq:updatey}. This validates our observation in Sec.~\ref{subsec:2b}.

\subsection{Inference Complexity and Training Complexity}
Finally, we evaluate the inference and training complexities of the learning-based methods, which determine the required computing resources in practical applications.

In Table \ref{table:time}, we list the inference complexities of the GNNs that can achieve at least 90\% of the SE of ``LISA-SDR'' in Fig.~\ref{fig:genK}, where the time complexity refers to the runtime for inference, the space complexity refers to the number of trainable weights, and the simulation setup in Fig.~\ref{fig:genK} is used. Since numerical algorithms can only run on CPU, the time complexity is obtained on an Intel Core i9-10940X CPU for a fair comparison. We can see that ``LISA-SDR'' requires a long time for convergence, whereas NGNN and SGNN are 50 $\sim$ 100 times faster. The storage demand of the trainable weights in a GNN is about 2.4~M bytes if single-precision floating-point format is used.

\begin{table}[htb!]
    \setlength\tabcolsep{2pt}
\centering
\vspace{-0mm}
\caption{Inference Complexity ($M\!=\!16, N_{\text{RF}}\!=\!6, N_T\!=\!16, N_R\!=\!2$)}\label{table:time}
\vspace{-2mm}
\footnotesize
    \begin{tabular}{c|c|c|c|c|c}
        \hline\hline
           & $K$& NGNN & SGNN & SGNN w/o F & LISA-SDR \\ \hline
           \multirow{3}{*}{\makecell{{\bf Time complexity}\\ (ms)}}& 20 & 81 & 101 & 98 & 8927 \\ \cline{2-6}
           & 40 & 98 & 155 & 150 & 9058 \\ \cline{2-6}
           & 60 & 112 & 179 & 173 & 9205  \\ \hline
          \bf Space complexity &-& 662k & 635k & 633k & -  \\ \hline
          \hline
    \end{tabular}
    \vspace{-0mm}
\end{table}

\begin{figure}[htb!]
	\centering
	\vspace{-0mm}
	\includegraphics[width=0.76\linewidth]{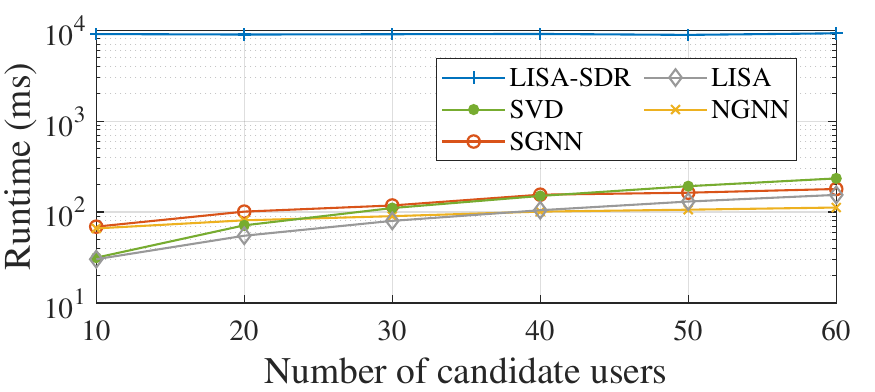}
	\vspace{-4mm}
	\caption{Runtime versus $K$, $M=16, N_{\text{RF}}=6, N_T=16, N_R=2$.}
        \label{fig:timeK}
	\vspace{-2mm}
\end{figure}

In Fig. \ref{fig:timeK}, we depict the runtime of different methods. It can be seen that the runtimes of ``LISA'' and ``SVD'' are comparable to those of the GNNs, yet their performance is inferior as demonstrated in Figs. \ref{fig:SNR}$\sim$\ref{fig:genNR}.

\begin{figure}[htb!]
	\centering
	\vspace{-0mm}
	\includegraphics[width=0.76\linewidth]{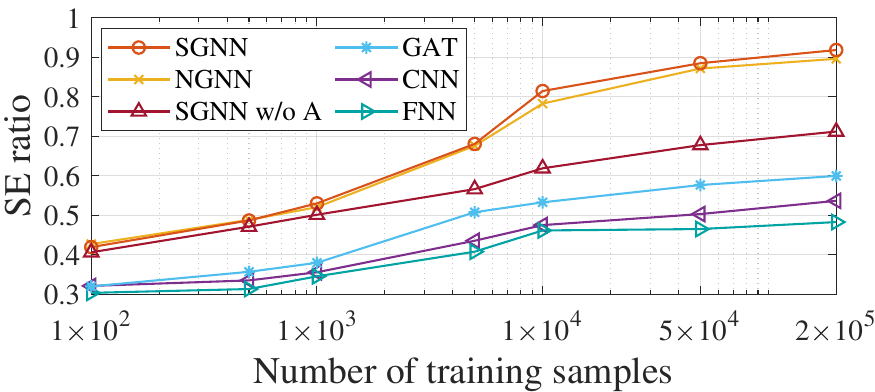}
	\vspace{-4mm}
	\caption{Impact of the number of training samples, $M=16, K=20, N_{\text{RF}}=4, N_T=16, N_R=2$.}
        \label{fig:sample}
	\vspace{-2mm}
\end{figure}

In Fig. \ref{fig:sample}, we provide the SE ratio achieved by the DNNs trained with different numbers of samples. It is shown that SGNN and NGNN outperform other DNNs.
CNN and FNN allocate the majority of power to a single user on an RB, whose performance grows slowly. This also explains why their performance hardly increases with $N_{\rm RF}$ in Fig.~\ref{fig:NRF}. GAT surpasses CNN and FNN, but the gap between GAT and the proposed methods is significant.

In Table \ref{table:train}, we provide the training complexities of different DNNs, where the two values before and after the slash are respectively for the scheduler and precoder modules.
The space, sample, and time complexities indicate the minimal requirements of trainable weights, training samples, and training time to achieve an expected performance.
To enable all DNNs to achieve the same performance, a narrow-band system is considered, and the performance is set as 85\% SE of ``LISA-SDR''. The DNNs are trained on NVIDIA GeForce RTX 3080 GPU. It is observed that the training complexities of NGNN and SGNN are lower than other DNNs, owing to exploiting permutation priors and introducing the attention mechanism.

\begin{table}[htb!]
    \setlength\tabcolsep{0.2pt}
\centering
\vspace{-0mm}
\caption{Training Complexity ($M=1, K=10, N_{\text{RF}}=4, N_T=16, N_R=1,  P_{tot}=40{\rm dBm}$)}\label{table:train}
\vspace{-2mm}
\footnotesize
\begin{threeparttable}
    \begin{tabular}{c|c|c|c|c|c}
        \hline\hline
          \bf Methods & NGNN & SGNN & GAT & CNN & FNN \\ \hline
          \bf Number of layers & 6/6 & 4/6 & 4/6 & 6/6 & 6/8 \\ \hline
          \bf Elements in hidden representation & 32/128 & 16/128 & 64/256 & 128/256 & 1024/2048 \\ \hline
          \bf Space complexity & 374k & 369k & 2.26M & 5.73M & 25.1M \\ \hline
          \bf Sample complexity & 4.0k & 3.3k & 380k & 900k & 1.4M \\ \hline
          {\bf Time complexity} (hour) & 0.19 & 0.24 & 2.0 & 12.3 & 4.6 \\ \hline
          \hline
    \end{tabular}
    \vspace{-0mm}
\begin{tablenotes}
 \footnotesize
        \item The elements in hidden representations are often referred to as ``channels'' in CNNs and ``neurons'' in FNNs.
\end{tablenotes}
\end{threeparttable}
\vspace{-2mm}
\end{table}

\section{Conclusion}\label{sec:con}
In this work, we proposed a GNN-based architecture to jointly optimize spatial-frequency user scheduling, hybrid precoding, and analog combining in MU-MIMO OFDM systems. The architecture consists of two cascaded modules dedicated to learning the discrete scheduling policy and the hybrid precoding policy, enabling joint training of the two modules to enhance overall learning performance. We discovered a same-parameter same-decision (SPSD) property for the wireless policies defined on sets. We found that a GNN cannot well learn the non-SPSD scheduling policy for users with similar channels, and proposed a sequence of GNNs for the scheduler module to improve the performance in high-density user scenarios. We noticed that the precoding policy is non-SPSD on the user set, hindering the generalizability of GNNs with linear aggregators to the number of users. We then proposed a novel attention mechanism for the precoder module to better aggregate information from different multi-antenna users, which can be generalized to all sizes.

Simulation results showcased several advantages of the proposed architecture in practical systems. The architecture can achieve SE comparable to the best of existing numerical algorithms but at remarkably shorter computing latency. The architecture can be well generalized to the numbers of users, RBs, and antennas at both the BS and users, avoiding the retraining for different system scales. The architecture requires much lower training complexity than other DNNs to achieve the same performance, facilitating fast adaptation to dynamic channel environments.

\begin{appendices}
\numberwithin{equation}{section}

\section{Proof of Example 1}\label{app:A}
{\bf On antenna set:} The optimal precoder matrix consists of ${\bf w}_k = \sqrt{p_k}\frac{({\bf I}_{N_T}+\sum_{l=1}^K {\bf h}_l{\bf h}_l^H\lambda_l/\sigma^2)^{-1}{\bf h}_k}{||({\bf I}_{N_T}+\sum_{l=1}^K {\bf h}_l{\bf h}_l^H\lambda_l/\sigma^2)^{-1}{\bf h}_k||_2} \in \mathbb{C}^{N_T\times1}, k = 1,\cdots,K$ \cite{optimalStructure}, where ${\bf h}_k$ is the channel vector of the $k$th user, $p_k$ and $\lambda_k$ are parameters that need to be further optimized. We next show that if $({\bf h}_k)_i = ({\bf h}_k)_j$ for $\forall k$ (i.e., $\tilde{\bf h}_i=\tilde{\bf h}_j$), then $({\bf w}_k)_i = ({\bf w}_k)_j$ for $\forall k$ (i.e., $\tilde{\bf w}_i=\tilde{\bf w}_j$), where $i,j\in \{1,\cdots,N_T\}$.

Denote ${\bf U} = [{\bf h}_1\sqrt{\lambda_1} / \sigma, \cdots, {\bf h}_K\sqrt{\lambda_K} / \sigma]$, and ${\bf w}'_k \triangleq ({\bf I}_{N_T} + \sum_{l=1}^K {\bf h}_l {\bf h}_l^H \lambda_l / \sigma^2)^{-1} {\bf h}_k = ({\bf I}_{N_T} + {\bf U}{\bf U}^H )^{-1}{\bf h}_k$. Using the Woodbury identity, we have ${\bf w}'_k = [{\bf I}_{N_T} - {\bf U}({\bf I}_K+{\bf U}^H{\bf U})^{-1}{\bf U}^H]{\bf h}_k = {\bf h}_k - {\bf U}({\bf I}_K+{\bf U}^H{\bf U})^{-1}{\bf U}^H{\bf h}_k = {\bf h}_k - {\bf U}{\bf z}_k$. We can see that $({\bf w}'_k)_i = ({\bf w}'_k)_j$ because $({\bf h}_k)_i = ({\bf h}_k)_j$ and the $i$th and $j$th rows are same in ${\bf U}$. Then, by multiplying with $\sqrt{p_k}/||{\bf w}'_k||_2$, we obtain $({\bf w}_k)_i = ({\bf w}_k)_j$. Therefore, the problem and the policy on antenna set is SPSD.

{\bf On user set:}
We prove that if ${\bf h}_i = {\bf h}_j = {\bf h}$, then ${\bf w}_i = {\bf w}_j = {\bf w}$ cannot be an optimal solution by a counter-example, where $i,j\in \{1,\cdots,K\}$. This can be verified by the fact that ${\bf w}_i = \sqrt{2}{\bf w}$ and ${\bf w}_j = 0$ (or ${\bf w}_i = 0$ and ${\bf w}_j = \sqrt{2}{\bf w}$) achieve higher sum rate than ${\bf w}_i = {\bf w}_j = {\bf w}$, since the sum rate achieved by the $i$th and $j$th users is $\log_2(1+\frac{|{\bf h}^H\sqrt{2}{\bf w}|^2}{\sum^K_{l\neq i,j}|{\bf h}^H{\bf w}_l|^2+\sigma^2})>2\log_2(1+\frac{|{\bf h}^H{\bf w}|^2}{|{\bf h}^H{\bf w}|^2+\sum^K_{l\neq i,j}|{\bf h}^H{\bf w}_l|^2+\sigma^2})$ meanwhile the sum rate of other users remains unchanged. Thus, the problem and the policy on user set is non-SPSD.

\section{Proof of Example 2}\label{app:B}
The power control problem can be expressed as $P_2: \max_{p_1,p_2} \log_2(1+\frac{h_{11}p_1}{h_{12}p_2+\sigma^2}) + \log_2(1+\frac{h_{22}p_2}{h_{21}p_1+\sigma^2}),~ {\rm s.t.}~ 0\leq p_1,p_2\leq P_{max}$, where $h_{11}$, $h_{22}$, $h_{12}$, and $h_{21}$ are channel gains and $P_{max}$ is the maximal transmit power of each transmitter.
The optimal solution of problem $P_2$ is one of $(p_1,p_2)=(P_{max},0)$, $(0,P_{max})$, and $(P_{max},P_{max})$ \cite{LZQ}, which can be obtained by comparing the SE achieved by the three combinations. Further considering the same-parameter assumption, which implies a symmetric channel characterized by $h_{11} = h_{22} = h_T$ and $h_{12}=h_{21} = h_I$, the optimal solution depends on a threshold $s_h = 2/(\sqrt{h_T^2+4h_I^2}-h_T)$. Specifically, the optimal solution is $(p_1,p_2)=(P_{max},P_{max})$ when $s_h>P_{max}/\sigma^2$, the optimal solutions are $(P_{max},0)$ and $(0,P_{max})$ when $s_h<P_{max}/\sigma^2$, and are all the three combinations when $s_h=P_{max}/\sigma^2$.

\end{appendices}

\bibliography{IEEEabrv,ref}

\begin{thebibliography}{10}
\providecommand{\url}[1]{#1}
\csname url@samestyle\endcsname
\providecommand{\newblock}{\relax}
\providecommand{\bibinfo}[2]{#2}
\providecommand{\BIBentrySTDinterwordspacing}{\spaceskip=0pt\relax}
\providecommand{\BIBentryALTinterwordstretchfactor}{4}
\providecommand{\BIBentryALTinterwordspacing}{\spaceskip=\fontdimen2\font plus
\BIBentryALTinterwordstretchfactor\fontdimen3\font minus \fontdimen4\font\relax}
\providecommand{\BIBforeignlanguage}[2]{{%
\expandafter\ifx\csname l@#1\endcsname\relax
\typeout{** WARNING: IEEEtran.bst: No hyphenation pattern has been}%
\typeout{** loaded for the language `#1'. Using the pattern for}%
\typeout{** the default language instead.}%
\else
\language=\csname l@#1\endcsname
\fi
#2}}
\providecommand{\BIBdecl}{\relax}
\BIBdecl

\bibitem{LSJ_WCNC}
S.~Liu and C.~Yang, ``Learning user scheduling and hybrid precoding with sequential graph neural network,'' in \emph{Proc. IEEE Wireless Commun. Netw. Conf. (WCNC)}, Dubai, United Arab Emirates, Apr. 2024, pp. 1--6.

\bibitem{overviewHY}
R.~W. Heath, N.~González-Prelcic, S.~Rangan, W.~Roh, and A.~M. Sayeed, ``An overview of signal processing techniques for millimeter wave {MIMO} systems,'' \emph{IEEE J. Sel. Top. Signal Process.}, vol.~10, no.~3, pp. 436--453, Apr. 2016.

\bibitem{MO}
X.~Yu, J.-C. Shen, J.~Zhang, and K.~B. Letaief, ``Alternating minimization algorithms for hybrid precoding in millimeter wave {MIMO} systems,'' \emph{IEEE J. Sel. Top. Signal Process.}, vol.~10, no.~3, pp. 485--500, Apr. 2016.

\bibitem{tmlcn}
Q.~An, S.~Segarra, C.~Dick, A.~Sabharwal, and R.~Doost-Mohammady, ``A deep reinforcement learning-based resource scheduler for massive {MIMO} networks,'' \emph{IEEE Trans. Mach. Learn. Commun. Netw.}, vol.~1, pp. 242--257, 2023.

\bibitem{LISA}
J.~P. González-Coma, W.~Utschick, and L.~Castedo, ``Hybrid {LISA} for wideband multiuser millimeter-wave communication systems under beam squint,'' \emph{IEEE Trans. Wireless Commun.}, vol.~18, no.~2, pp. 1277--1288, Feb. 2019.

\bibitem{Bogale}
T.~E. Bogale, L.~B. Le, A.~Haghighat, and L.~Vandendorpe, ``On the number of {RF} chains and phase shifters, and scheduling design with hybrid analog-digital beamforming,'' \emph{IEEE Trans. Wireless Commun.}, vol.~15, no.~5, pp. 3311--3326, May 2016.

\bibitem{KLX}
L.~Kong, S.~Han, and C.~Yang, ``Wideband hybrid precoder for massive {MIMO} systems,'' in \emph{Proc. IEEE Global Conf. Signal Inf. Process. (GlobalSIP)}, Orlando, FL, USA, Dec. 2015, pp. 305--309.

\bibitem{FNNsupervised}
M.~Mohammadkarimi, M.~Darabi, and B.~Maham, ``User scheduling in massive {MIMO}: A joint deep learning and genetic algorithm approach,'' in \emph{Proc. IEEE 95th Veh. Technol. Conf. (VTC-Spring)}, Helsinki, Finland, Jun. 2022, pp. 1--6.

\bibitem{SPAWCsupervised}
B.~Song, Z.~Zhou, C.~Li, D.~Guo, X.~Fu, and M.~Hong, ``Transformer based approach for wireless resource allocation problems involving mixed discrete and continuous variables,'' in \emph{Proc. IEEE Workshop Signal Process. Adv. Wireless Commun. (SPAWC)}, Shanghai, China, Sep. 2023, pp. 636--640.

\bibitem{RL-SLNR}
I.~M. Braga, E.~d.~O. Cavalcante, G.~Fodor, Y.~C.~B. Silva, C.~F.~M. e~Silva, and W.~C. Freitas, ``User scheduling based on multi-agent deep {Q}-learning for robust beamforming in multicell {MISO} systems,'' \emph{IEEE Commun. Lett.}, vol.~24, no.~12, pp. 2809--2813, Dec. 2020.

\bibitem{RLschedule2}
W.~V.~F. Mauricio, T.~F. Maciel, A.~Klein, and F.~R.~M. Lima, ``Scheduling for massive {MIMO} with hybrid precoding using contextual multi-armed bandits,'' \emph{IEEE Trans. Veh. Technol.}, vol.~71, no.~7, pp. 7397--7413, Jul. 2022.

\bibitem{transformer}
A.~Vaswani, N.~Shazeer, N.~Parmar, J.~Uszkoreit, L.~Jones, A.~N. Gomez, {\L}.~Kaiser, and I.~Polosukhin, ``Attention is all you need,'' in \emph{Proc. Adv. Neural Inf. Process. Syst. (NeurIPS)}, Long Beach, CA, USA, 2017.

\bibitem{codebook}
Y.~Yan, B.~Zhang, C.~Li, J.~Bai, and Z.~Yao, ``A novel model-assisted decentralized multi-agent reinforcement learning for joint optimization of hybrid beamforming in massive {MIMO} {mmWave} systems,'' \emph{IEEE Trans. Veh. Technol.}, vol.~72, no.~11, pp. 14\,743--14\,755, Nov. 2023.

\bibitem{yuweiSchedule}
Z.~Zhang, T.~Jiang, and W.~Yu, ``Learning based user scheduling in reconfigurable intelligent surface assisted multiuser downlink,'' \emph{IEEE J. Sel. Top. Signal Process.}, vol.~16, no.~5, pp. 1026--1039, Aug. 2022.

\bibitem{RLschedule1}
R.~Huang and V.~W.~S. Wong, ``Joint user scheduling, phase shift control, and beamforming optimization in intelligent reflecting surface-aided systems,'' \emph{IEEE Trans. Wireless Commun.}, vol.~21, no.~9, pp. 7521--7535, Sep. 2022.

\bibitem{HSW}
S.~He, J.~Yuan, Z.~An, W.~Huang, Y.~Huang, and Y.~Zhang, ``Joint user scheduling and beamforming design for multiuser {MISO} downlink systems,'' \emph{IEEE Trans. Wireless Commun.}, vol.~22, no.~5, pp. 2975--2988, May 2023.

\bibitem{LYreview}
M.~Lee, G.~Yu, H.~Dai, and G.~Y. Li, ``Graph neural networks meet wireless communications: Motivation, applications, and future directions,'' \emph{IEEE Wireless Commun.}, vol.~29, no.~5, pp. 12--19, Oct. 2022.

\bibitem{modelGNN}
J.~Guo and C.~Yang, ``A model-based {GNN} for learning precoding,'' \emph{IEEE Trans. Wireless Commun.}, vol.~23, no.~7, pp. 6983--6999, Jul. 2024.

\bibitem{LSJ_TWC}
S.~Liu, J.~Guo, and C.~Yang, ``Multidimensional graph neural networks for wireless communications,'' \emph{IEEE Trans. Wireless Commun.}, vol.~23, no.~4, pp. 3057--3073, Apr. 2024.

\bibitem{ZBC_TCOM}
B.~Zhao, J.~Guo, and C.~Yang, ``Understanding the performance of learning precoding policies with graph and convolutional neural networks,'' \emph{IEEE Trans. Commun.}, vol.~72, no.~9, pp. 5657--5673, Sep. 2024.

\bibitem{dll}
H.~Jiang, Y.~Lu, X.~Li, B.~Wang, Y.~Zhou, and L.~Dai, ``Attention-based hybrid precoding for {mmWave} {MIMO} systems,'' in \emph{Proc. IEEE Inf. Theory Workshop (ITW)}, Kanazawa, Japan, Oct. 2021, pp. 1--6.

\bibitem{GATbeijiao}
Y.~Li, Y.~Lu, B.~Ai, O.~A. Dobre, Z.~Ding, and D.~Niyato, ``{GNN}-based beamforming for sum-rate maximization in {MU-MISO} networks,'' \emph{IEEE Trans. Wireless Commun.}, vol.~23, no.~8, pp. 9251--9264, Aug. 2024.

\bibitem{GATbeijiao2}
Y.~Li, Y.~Lu, R.~Zhang, B.~Ai, and Z.~Zhong, ``Deep learning for energy efficient beamforming in {MU-MISO} networks: A {GAT}-based approach,'' \emph{IEEE Wireless Commun. Lett.}, vol.~12, no.~7, pp. 1264--1268, Jul. 2023.

\bibitem{GAT}
P.~Veli{\v{c}}kovi{\'c}, G.~Cucurull, A.~Casanova, A.~Romero, P.~Lio, and Y.~Bengio, ``Graph attention networks,'' in \emph{Proc. 6th Int. Conf. Learn. Represent. (ICLR)}, Vancouver, BC, Canada, 2018, pp. 2920--2931.

\bibitem{liyang}
Y.~Li and Y.-F. Liu, ``{HPE} transformer: Learning to optimize multi-group multicast beamforming under nonconvex {QoS} constraints,'' \emph{IEEE Trans. Commun.}, vol.~72, no.~9, pp. 5581--5594, Sep. 2024.

\bibitem{RGNN}
J.~Guo and C.~Yang, ``Recursive {GNNs} for learning precoding policies with size-generalizability,'' \emph{IEEE Trans. Mach. Learn. Commun. Netw.}, vol.~2, pp. 1558--1579, 2024.

\bibitem{SVDYH}
H.~Yuan, J.~An, N.~Yang, K.~Yang, and T.~Q. Duong, ``Low complexity hybrid precoding for multiuser millimeter wave systems over frequency selective channels,'' \emph{IEEE Trans. Veh. Technol.}, vol.~68, no.~1, pp. 983--987, Jan. 2019.

\bibitem{Sun2019Learning}
C.~Sun and C.~Yang, ``Learning to optimize with unsupervised learning: Training deep neural networks for {URLLC},'' in \emph{Proc. IEEE 30th Int. Symp. Pers. Indoor Mob. Radio Commun. (PIMRC)}, Istanbul, Turkey, Sep. 2019, pp. 1--7.

\bibitem{MSUS}
E.~Conte, S.~Tomasin, and N.~Benvenuto, ``A simplified greedy algorithm for joint scheduling and beamforming in multiuser {MIMO OFDM},'' \emph{IEEE Commun. Lett.}, vol.~14, no.~5, pp. 381--383, May 2010.

\bibitem{SUS}
T.~Yoo and A.~Goldsmith, ``On the optimality of multiantenna broadcast scheduling using zero-forcing beamforming,'' \emph{IEEE J. Sel. Areas Commun.}, vol.~24, no.~3, pp. 528--541, Mar. 2006.

\bibitem{top-k}
S.~M. Xie and S.~Ermon, ``Reparameterizable subset sampling via continuous relaxations,'' in \emph{Proc. 28th Int. Joint Conf. Artif. Intell. (IJCAI)}, Macao, China, 2019, pp. 3919--3925.

\bibitem{MOcomplexity}
X.~Zhao, T.~Lin, Y.~Zhu, and J.~Zhang, ``Partially-connected hybrid beamforming for spectral efficiency maximization via a weighted {MMSE} equivalence,'' \emph{IEEE Trans. Wireless Commun.}, vol.~20, no.~12, pp. 8218--8232, Dec. 2021.

\bibitem{EIGLY}
Y.~Liu and J.~Wang, ``Low-complexity {OFDM}-based hybrid precoding for multiuser massive {MIMO} systems,'' \emph{IEEE Wireless Commun. Lett.}, vol.~9, no.~3, pp. 263--266, Mar. 2020.

\bibitem{OMP}
O.~E. Ayach, S.~Rajagopal, S.~Abu-Surra, Z.~Pi, and R.~W. Heath, ``Spatially sparse precoding in millimeter wave {MIMO} systems,'' \emph{IEEE Trans. Wireless Commun.}, vol.~13, no.~3, pp. 1499--1513, Mar. 2014.

\bibitem{3gpp38901}
3GPP, ``Study on channel model for frequencies from 0.5 to 100 {GHz},'' TR 38.901, Version 17.0.0, Mar. 2022.

\bibitem{CNNprecode}
A.~M. Elbir, K.~V. Mishra, M.~R.~B. Shankar, and B.~Ottersten, ``A family of deep learning architectures for channel estimation and hybrid beamforming in multi-carrier {mm-Wave} massive {MIMO},'' \emph{IEEE Trans. Cogn. Commun. Netw.}, vol.~8, no.~2, pp. 642--656, Jun. 2022.

\bibitem{optimalStructure}
E.~Björnson, M.~Bengtsson, and B.~Ottersten, ``Optimal multiuser transmit beamforming: A difficult problem with a simple solution structure [lecture notes],'' \emph{IEEE Signal Process. Mag.}, vol.~31, no.~4, pp. 142--148, Jul. 2014.

\bibitem{LZQ}
Z.-Q. Luo and S.~Zhang, ``Dynamic spectrum management: Complexity and duality,'' \emph{IEEE J. Sel. Top. Signal Process.}, vol.~2, no.~1, pp. 57--73, Feb. 2008.

\end{thebibliography}

\end{document}